\documentclass[
 reprint,
 amsmath,
 amssymb,
 aps,
 pra,
 floatfix,
 longbibliography
]{revtex4-2}

\usepackage[final]{graphicx}
\usepackage{booktabs}
\usepackage{bm}
\usepackage{microtype}
\usepackage[hidelinks]{hyperref}

\hypersetup{
 pdftitle={Exact Haar Statistics of Planar k-Purity in Multipartite Quantum Systems},
 pdfauthor={Kourosh Mirsohi},
 pdfsubject={Haar-random entanglement statistics and planar k-uniform states},
 pdfkeywords={planar k-uniform states, Haar random states, subsystem purity, multipartite entanglement}
}

\newcommand{\ket}[1]{\lvert #1\rangle}
\newcommand{\bra}[1]{\langle #1\rvert}
\newcommand{\proj}[1]{\ket{#1}\!\bra{#1}}
\newcommand{\Tr}{\operatorname{Tr}}
\newcommand{\E}{\mathbb{E}}
\newcommand{\Var}{\operatorname{Var}}
\newcommand{\Pplanark}{\pi_{\mathrm P}^{(k)}}
\newcommand{\Pplanar}{\pi_{\mathrm P}^{(r)}}
\newcommand{\Pabsolute}{\pi_{\mathrm A}}
\newcommand{\rising}[2]{(#1)_{#2}}

\begin{document}

\title{Exact Haar Statistics of Planar \texorpdfstring{$k$}{k}-Purity in Multipartite Quantum Systems}

\author{Kourosh Mirsohi}
\email{mirsohi@gmail.com}
\affiliation{Department of Computer Science, University of California, Irvine, California 92697, USA}

\date{August 2, 2026}

\begin{abstract}
Planar $k$-uniform states have maximally mixed reductions on every $k$-site
interval of a ring.  We introduce planar $k$-purity---the mean purity of those
intervals---as a faithful cost function for this simultaneous constraint.  For
a Haar-random pure state of $n$ parties with local dimension $p$, the joint
purity of two intervals depends only on their overlap.  An exhaustive
four-replica calculation therefore yields the complete cyclic covariance kernel
and a closed variance for every $1\le k\le\lfloor n/2\rfloor$.  In the balanced
qubit case, $k=\lfloor n/2\rfloor$ and $N=2^n$, the variance is asymptotic to
$20/(3nN^2)$ for even $n$ and $6/(nN^2)$ for odd $n$.  A site-factorized
permutation representation gives all raw moments and permits exact finite-sum
evaluation of the skewness.  Fixed-seed simulations validate nonbalanced qubit
and qutrit cases, while balanced-qubit simulations through $n=10$ validate the
parity formulas and finite-size skewness.  Balanced planar and absolute
balanced purity have the same Haar mean but different fluctuations; the planar
variance is approximately $2.83$ times larger at $n=10$.  Thus subsystem
incidence, although invisible to every one-cut marginal distribution, controls
the collective fluctuations of geometrically related cuts.  Together with
universal attainability and a linear number of interval constraints at
balance, this provides a geometry-aware benchmark for multipartite
entanglement beyond AME existence regimes.
\end{abstract}

\maketitle

\section{Introduction}
\label{sec:introduction}

Entanglement in a many-body pure state can be nearly maximal across one
bipartition and highly structured across another.  This tension motivates two
complementary questions.  A typicality question asks what most states drawn
from a natural ensemble look like.  An extremal question asks whether one
state can satisfy simultaneous maximal-entanglement constraints across a
prescribed family of cuts.  Haar-random states provide the canonical setting
for the first question; uniformity conditions specify different versions of
the second by choosing which reduced states must be maximally mixed.

For a bipartite pure state $\ket{\psi}_{AB}$, the reduced-state purity
\begin{equation}
 \pi_A(\psi)=\Tr\!\left(\rho_A^2\right),
 \qquad
 \rho_A=\Tr_B\proj{\psi},
 \label{eq:subsystem-purity}
\end{equation}
ranges from $1/d_A$ to $1$ when $d_A\le d_B$.  The minimum occurs exactly
when $\rho_A=I_A/d_A$, so lower purity means greater entanglement across the
specified cut.  Purity is the exponential of the negative second R\'{e}nyi
entropy, $\pi_A=e^{-S_2(A)}$, and is technically convenient because it is a
polynomial of degree four in the state amplitudes.  It is not, by itself, a
complete measure of multipartite entanglement.  A family or average of
subsystem purities nevertheless records how entanglement is distributed over
a selected set of bipartitions~\cite{Horodecki2009,Facchi2006,Facchi2008}.

The typicality side of the problem is well established for one fixed
bipartition.  The induced reduced state of a Haar-random vector is a normalized
Wishart matrix~\cite{Lubkin1978,Zyczkowski2001}.  Its average purity has a
simple rational form, and its entropy is close to maximal when the environment
is sufficiently large~\cite{Page1993,Hayden2006}.  Exact purity moments and
finite-dimensional distributions have also been derived
~\cite{Giraud2007Moments,Giraud2007Distribution}.  These results underlie
canonical typicality: a small subsystem of a generic global pure state often
resembles a thermal or maximally mixed state even though the global evolution
is pure~\cite{Goldstein2006,Popescu2006}.

The multipartite extremal problem is more restrictive.  An absolutely
maximally entangled (AME) state is maximally entangled across every cut with no
more than half of the parties on one side.  Equivalently, each such subsystem
is maximally mixed.  AME states connect multiunitary matrices and combinatorial
designs to quantum error-correcting codes, teleportation, and secret
sharing~\cite{Scott2004,Helwig2012,Helwig2013,Goyeneche2015,Rajchel2025}.
They can also be regarded as perfect tensors, which are basic ingredients in
tensor-network models of holographic error correction
~\cite{Pastawski2015,Geng2022,Gross2025}.  The simultaneous constraints are
frustrated: AME states do not exist for every party number and local dimension;
for example, no seven-qubit AME state exists~\cite{Huber2017}.

Planar $k$-uniformity instead requires maximal mixing only for every $k$-party
contiguous block on a circle~\cite{Wang2021}.  Geometry therefore becomes part
of the constraint, and $k$ controls its scale.  Wang's construction establishes
minimal-support planar $k$-uniform states for every $p\ge2$ and $n\ge2k$ (with
the $k=1$ case supplied by the generalized GHZ construction), so the defining
lower bound is attainable throughout the parameter range studied here.  An AME
state is planar $k$-uniform for every admissible $k$, whereas the converse need
not hold: even at balance, planar uniformity constrains only the cyclic
intervals.  This intermediate notion separates ordered subsystem constraints
from permutation-independent multipartite uniformity.  The same distinction
appears in tensor networks, where perfect tensors satisfy all balanced
isometry conditions while block-perfect conditions retain only cuts compatible
with an ordering of the tensor legs~\cite{Steinberg2024}.

This restriction is operationally meaningful rather than merely terminological.
Every contiguous $k$-site block of a planar $k$-uniform state is maximally
entangled with its complement, so communication and tensor-isometry protocols
that use such a cut remain available when the participating groups respect the
cyclic ordering.  At the same time, planar uniformity imposes only $n$ local
constraints instead of all $\binom nk$ size-$k$ constraints, and unlike AME
states it exists for every admissible $n,k,p$.  These features make planar
uniformity a natural target for ordered quantum architectures and motivate a
statistical baseline for its purity cost.

\subsection{Problem and contributions}
\label{sec:context}

A scalar ``multipartite purity'' is defined only after specifying the family of
cuts being averaged.  The statistical-mechanics approach to multipartite
entanglement uses all balanced bipartitions
~\cite{Facchi2008,Facchi2009,Facchi2010Frustration,Facchi2010}; recent work also
compares those statistics across Haar, Hadamard, and hypergraph-state
ensembles~\cite{Trotta2026}.  Planar $k$-purity instead uses the cyclic family
of $k$-site intervals.  Although a fixed interval has the same Haar marginal
distribution as any subsystem of the same size, the translated intervals are
correlated through their intersections.  Their variance therefore cannot be
obtained by treating the purities as independent samples.

The two-subsystem Haar moment and its organization by intersection size are
established ingredients of the absolute balanced-purity calculation
~\cite{Facchi2010}.  The contribution here is to solve the resulting incidence
problem for cyclic intervals.  Their pair multiplicities reduce to an explicit
displacement profile, yielding arbitrary-$k$, arbitrary-$p$ closed sums, the
complete spatial covariance kernel, and compact parity-dependent balanced
formulas.  Averaging the same kernel with binomial all-subset multiplicities
recovers the known absolute-purity variance and provides an independent
analytic check.

The calculation combines the swap representation of purity, the Haar projector
on replicated states, Kronecker-delta equality graphs, and finite geometric
sums over interval displacements.  We derive the exact mean, two-interval
moment, covariance kernel, and variance for arbitrary local dimension and every
$1\le k\le\lfloor n/2\rfloor$; identify the fixed-$k$, extensive, and balanced
asymptotic regimes; and give a site-factorized exact representation of all raw
moments.  Numerically evaluated six-replica sums provide finite-size skewness,
and independent simulations test the general-$k$, qutrit, and balanced-qubit
predictions with explicit uncertainty estimates.  Because the method depends
on subsystem incidence, it extends naturally to other specified subsystem
families.  Section~\ref{sec:discussion} separates the statistical conclusions
from the distinct questions of state preparation and rare-event sampling.

\section{Planar \texorpdfstring{$k$}{k}-purity and the balanced comparison}
\label{sec:functionals}

\begin{figure}[tbp]
 \centering
 \includegraphics[width=0.76\columnwidth]{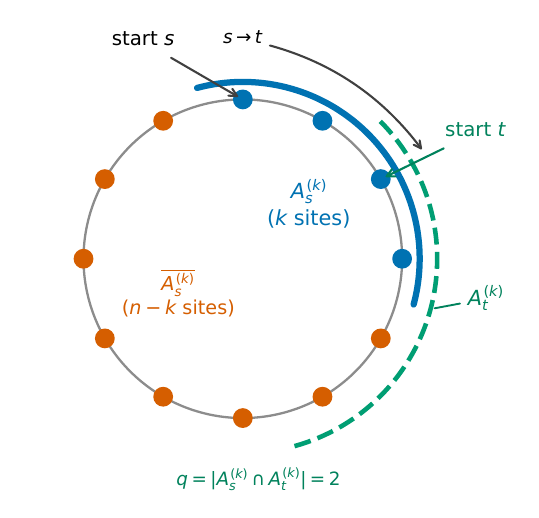}
 \caption{Subsystem geometry for planar $k$-purity.  The blue sites form one
 $k$-site interval $A_s^{(k)}$; the orange sites form its $(n-k)$-site
 complement.  The arrow fixes the translation orientation, and the dashed
 green arc shows a translated interval $A_t^{(k)}$ with overlap
 $q=|A_s^{(k)}\cap A_t^{(k)}|$.  Translating through all starts in
 $\mathbb Z_n$ produces the reductions averaged in
 Eq.~\eqref{eq:planar-purity}.}
 \label{fig:partition}
\end{figure}

Consider $n$ parties of local dimension $p\ge2$ on a ring.  Let
\begin{equation}
 r=\left\lfloor\frac n2\right\rfloor,
 \qquad N=p^n,
 \qquad 1\le k\le r,
\end{equation}
and label the sites by $\mathbb Z_n$.  For each $s\in\mathbb Z_n$, define
\begin{equation}
 A_s^{(k)}=\{s,s+1,\ldots,s+k-1\}\pmod n.
 \label{eq:cyclic-block}
\end{equation}
Figure~\ref{fig:partition} illustrates the associated bipartition.

We define the planar $k$-purity by
\begin{equation}
 \Pplanark(\psi)=\frac1n\sum_{s=0}^{n-1}\pi_{A_s^{(k)}}(\psi).
 \label{eq:planar-purity}
\end{equation}
Each term is bounded below by $p^{-k}$, and therefore
\begin{equation}
 \Pplanark(\psi)\ge p^{-k}.
 \label{eq:planar-lower-bound}
\end{equation}
Equality holds exactly when all $n$ reductions on adjacent $k$ parties are
maximally mixed: this is the planar $k$-uniform condition of
Ref.~\cite{Wang2021}.  The existence result recalled above makes $p^{-k}$ the
attained minimum for all admissible $n,k,p$.  Thus $\Pplanark-p^{-k}$ is a
faithful nonnegative cost function: it vanishes precisely on the planar
$k$-uniform minimizer set.  If equality holds, every smaller contiguous
reduction is also maximally mixed by partial tracing.

Our simulations and comparison with the absolute functional use the strongest
admissible interval size, $k=r$.  We abbreviate
$\Pplanar\equiv\pi_{\mathrm P}^{(r)}$.  When $n$ is even,
$A_{s+r}^{(r)}=(A_s^{(r)})^c$ and the two purities are identical for every
global pure state; Eq.~\eqref{eq:planar-purity} therefore equals the average
over the $n/2$ distinct unordered interval bipartitions.  When $n$ is odd, all
$n$ balanced interval cuts are distinct.  Retaining the $1/n$ form is the
parity-consistent convention used below.

For comparison, define the absolute purity
\begin{equation}
 \Pabsolute(\psi)=\frac1{|\mathcal B_r|}
 \sum_{A\in\mathcal B_r}\pi_A(\psi),
 \label{eq:absolute-purity}
\end{equation}
where $\mathcal B_r$ contains all $r$-site subsets when $n$ is odd and one
representative from each complementary pair when $n$ is even.  Its minimum is
attained by an AME state whenever such a state exists.

\section{Exact Haar moments}
\label{sec:moments}

Let $\ket\psi$ be Haar random in $\mathbb C^N$.  For a replica number $m\ge1$,
\begin{align}
 \E\!\left[(\proj\psi)^{\otimes m}\right]
 &=\frac1{\rising Nm}\sum_{\sigma\in S_m}U_\sigma,
 \label{eq:haar-moment}\\
 \rising Nm&=N(N+1)\cdots(N+m-1),
\end{align}
where $U_\sigma$ permutes the $m$ replicas according to
$\sigma$~\cite{Collins2006}.  Equation~\eqref{eq:haar-moment} is the normalized
projector onto the symmetric subspace.  It avoids coordinate-by-coordinate
phase counting and automatically handles repeated indices.  Throughout,
permutation products act from right to left:
$(\tau\sigma)(a)=\tau(\sigma(a))$.

\subsection{Mean}

Let $F_A$ swap subsystem $A$ between two replicas and act trivially on its complement.  The swap trick gives
\begin{equation}
 \pi_A(\psi)=\Tr\!\left[(\proj\psi)^{\otimes2}F_A\right].
\end{equation}
Using Eq.~\eqref{eq:haar-moment} with $m=2$, and writing
$a=p^k$ and $b=p^{n-k}$, we obtain
\begin{align}
 \E[\pi_A]
 &=\frac{\Tr(F_A)+\Tr(F_AF_{AB})}{N(N+1)}\\
 &=\frac{ab^2+a^2b}{N(N+1)}
 =\frac{a+b}{N+1}.
 \label{eq:mean}
\end{align}
The result is independent of the selected subsystem.  Hence
\begin{equation}
 \boxed{\E[\Pplanark]=\frac{p^k+p^{n-k}}{N+1}.}
 \label{eq:planar-mean}
\end{equation}
For the balanced choice $k=r$, every term in $\Pabsolute$ has the same
dimensions, so
\begin{equation}
 \boxed{\E[\Pplanar]=\E[\Pabsolute]
 =\frac{p^r+p^{n-r}}{N+1}.}
 \label{eq:shared-mean}
\end{equation}

\subsection{Second moment}

Consider two $k$-site blocks $A_s^{(k)}$ and $A_t^{(k)}$ and write
\begin{equation}
 q=|A_s^{(k)}\cap A_t^{(k)}|,
 \quad x=p^q,
 \quad y=p^{k-q},
 \quad w=p^{n-2k+q}.
 \label{eq:region-dimensions}
\end{equation}
The four disjoint regions $A_s^{(k)}\cap A_t^{(k)}$,
$A_s^{(k)}\setminus A_t^{(k)}$, $A_t^{(k)}\setminus A_s^{(k)}$, and
$(A_s^{(k)}\cup A_t^{(k)})^c$ have dimensions $x,y,y,w$, respectively.

Apply Eq.~\eqref{eq:haar-moment} with $m=4$ and insert the swaps $(12)$
on $A_s^{(k)}$ and $(34)$ on $A_t^{(k)}$.  For each $\sigma\in S_4$, the trace
factorizes over the four regions.  The 24 permutations reduce to the eight
contraction classes in Table~\ref{tab:contraction-classes}.  This exhaustive
enumeration supplies the pair kernel used here; the geometric step is its
evaluation over the cyclic interval incidence profile.

\begin{table}[b]
\caption{Contraction classes contributing to the two-purity moment.  The
representative uses the permutation convention stated below
Eq.~\eqref{eq:haar-moment}; the remaining columns give the multiplicity and
the powers in the factorized trace.}
\label{tab:contraction-classes}
\begin{ruledtabular}
\begin{tabular}{ccccc}
$\sigma$ & multiplicity & power of $x$ & power of $y$ & power of $w$\\
\hline
$(1324)$    & $2$ & $1$ & $4$ & $1$\\
$(23)$      & $4$ & $1$ & $4$ & $3$\\
$(13)(24)$  & $2$ & $2$ & $2$ & $2$\\
$(234)$     & $8$ & $2$ & $4$ & $2$\\
$e$         & $1$ & $2$ & $6$ & $4$\\
$(1234)$    & $4$ & $3$ & $4$ & $1$\\
$(34)$      & $2$ & $3$ & $6$ & $3$\\
$(12)(34)$  & $1$ & $4$ & $6$ & $2$
\end{tabular}
\end{ruledtabular}
\end{table}

Define the trace polynomial
\begin{align}
 T_{n,k}^{(p)}(q)={}&2xy^4w+4xy^4w^3+2x^2y^2w^2
 +8x^2y^4w^2\nonumber\\
 &+x^2y^6w^4+4x^3y^4w+2x^3y^6w^3+x^4y^6w^2.
 \label{eq:trace-polynomial}
\end{align}

\subsubsection{Kronecker-delta contraction picture}
\label{subsec:delta-picture}

The permutation trace has an equivalent coordinate representation that makes
the geometric content of Eq.~\eqref{eq:trace-polynomial} explicit.  For a
fixed bipartition $A\cup\bar A$, write
\begin{equation}
 \ket\psi=\sum_{\bm a,\bm b}z_{\bm a\bm b}
 \ket{\bm a}_A\ket{\bm b}_{\bar A}.
\end{equation}
Introducing four global computational-basis strings $\bm u_1,\ldots,\bm u_4$
gives
\begin{align}
 \pi_A={}&\sum_{\bm u_1,\ldots,\bm u_4}
 z_{\bm u_1}z_{\bm u_2}
 \bar z_{\bm u_3}\bar z_{\bm u_4}\,
 \Delta_A(\bm u_1,\bm u_2;\bm u_3,\bm u_4),
 \label{eq:purity-delta}\\
 \Delta_A={}&
 \delta_{\bm u_{1,A},\bm u_{4,A}}
 \delta_{\bm u_{2,A},\bm u_{3,A}}
 \delta_{\bm u_{1,\bar A},\bm u_{3,\bar A}}
 \delta_{\bm u_{2,\bar A},\bm u_{4,\bar A}}.
 \label{eq:delta-kernel}
\end{align}
Here $\delta_{\bm u,\bm v}$ is one only when the two restricted basis strings
are identical.  In the Haar average of $\pi_A\pi_{A'}$, the fourth-order
amplitude moment pairs the four unstarred amplitudes with the four starred
amplitudes in all $4!$ ways.  Thus the product initially contains eight
computational-basis strings, but after one Haar pairing is selected each
starred string is identified with one unstarred string.  The remaining sum is
a product of Kronecker deltas among four replica strings.

The sum can be performed independently at each site.  Draw one vertex for
each replica index and join vertices whose local labels are equated by a delta.  If
the resulting equality graph has $c$ connected components, summing the four
labels contributes $p^c$.  Only four local graphs occur, according to whether
the site lies in $A\cap A'$, $A\setminus A'$, $A'\setminus A$, or
$(A\cup A')^c$.  Their numbers of sites are $q$, $k-q$, $k-q$, and
$n-2k+q$, respectively.  Thus the delta-graph count is identical to the
cycle count in the replica formula.

\begin{figure}[tbp]
 \centering
 \includegraphics[width=0.94\columnwidth]{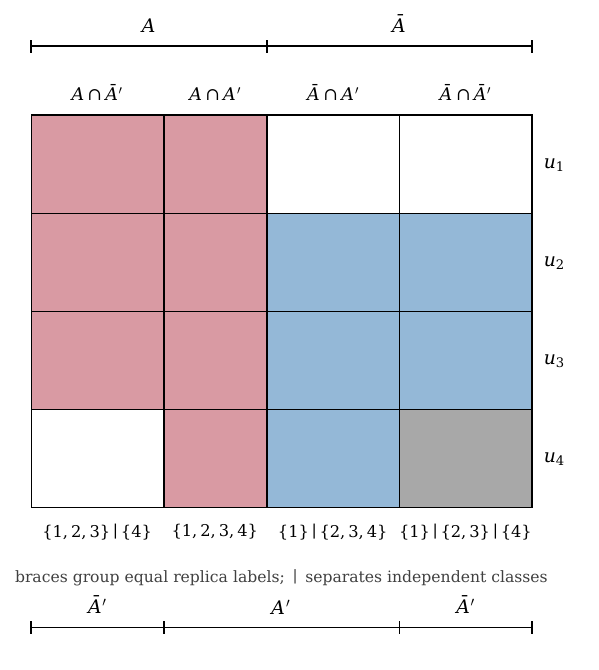}
 \caption{Representative Kronecker-delta contraction for $\sigma=(23)$.
 The four columns represent $A\setminus A'$, $A\cap A'$,
 $A'\setminus A$, and $(A\cup A')^c$, with dimensions $y,x,y,w$.
 Equal fills identify local replica labels; the braces give the same equality
 classes algebraically.  The component counts $(2,1,2,3)$ produce
 $xy^4w^3$, with multiplicity four in
 Table~\ref{tab:contraction-classes}.  Region widths are schematic.}
 \label{fig:delta-three}
\end{figure}

\begin{figure}[tbp]
 \centering
 \includegraphics[width=0.94\columnwidth]{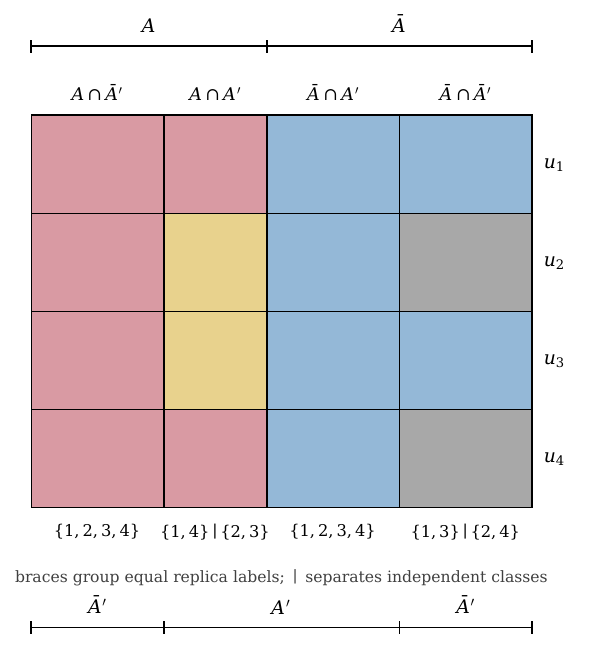}
 \caption{Crossed contraction for $\sigma=(13)(24)$, the third row of
 Table~\ref{tab:contraction-classes}.  The component counts $(1,2,1,2)$
 across regions of dimensions $y,x,y,w$ give $x^2y^2w^2$ with multiplicity
 two.  Because all four region dimensions depend on the translated overlap,
 this contraction displays the origin of the displacement-dependent
 covariance.}
 \label{fig:delta-crossed}
\end{figure}

In the labels below Figs.~\ref{fig:delta-three} and
\ref{fig:delta-crossed}, braces collect replicas whose local indices are
equal, while a vertical bar separates independent equality classes.  For
example, $\{1,4\}\mid\{2,3\}$ means $u_1=u_4$ and $u_2=u_3$, with no equality
imposed between the two pairs.  Consequently, the number of brace-delimited
classes is precisely the component count $c$ that determines the local factor
$p^c$.

Figure~\ref{fig:delta-three} illustrates a contraction whose equality
classes in the four displayed regions are
$\{1,2,3\}\mid\{4\}$, $\{1,2,3,4\}$,
$\{1\}\mid\{2,3,4\}$, and $\{1\}\mid\{2,3\}\mid\{4\}$.  It contributes
\begin{equation}
 y^2\,x\,y^2\,w^3=xy^4w^3,
\end{equation}
and four Haar pairings yield this monomial, as recorded in
Table~\ref{tab:contraction-classes}.

In the crossed contraction of Fig.~\ref{fig:delta-crossed}, the equality
classes are $\{1,2,3,4\}$, $\{1,4\}\mid\{2,3\}$,
$\{1,2,3,4\}$, and $\{1,3\}\mid\{2,4\}$.  The corresponding
factor is
\begin{equation}
 y\,x^2\,y\,w^2=x^2y^2w^2,
\end{equation}
with multiplicity two.  Unlike contractions whose component count is
independent of the cut placement, this term changes with
$q=|A\cap A'|$.  Translating $A'$ around the ring changes the widths of the
four regions, and the displacement sum becomes the finite geometric sums
evaluated below and in Appendix~\ref{app:second-moment}.

Then
\begin{equation}
 \E[\pi_{A_s^{(k)}}\pi_{A_t^{(k)}}]
 =\frac{T_{n,k}^{(p)}(q)}{\rising N4}.
 \label{eq:two-purity-moment}
\end{equation}

For relative displacement $d=t-s\pmod n$, the overlap is
\begin{equation}
 q_d^{(k)}=\max(0,k-d)+\max(0,k-n+d).
 \label{eq:overlap}
\end{equation}
This expression also gives the full spatial covariance kernel.  If
$C_{n,k}^{(p)}(d)=\operatorname{Cov}
(\pi_{A_0^{(k)}},\pi_{A_d^{(k)}})$, then
\begin{align}
 C_{n,k}^{(p)}(d)&=\frac{T_{n,k}^{(p)}(q_d^{(k)})}{\rising N4}
 -\left(\frac{a+b}{N+1}\right)^2,\nonumber\\
 \Var(\Pplanark)&=\frac1n\sum_{d=0}^{n-1}C_{n,k}^{(p)}(d).
 \label{eq:covariance-kernel}
\end{align}
There are $n$ ordered pairs $(s,t)$ for each $d$, so
\begin{equation}
 \E[(\Pplanark)^2]
 =\frac1{n\rising N4}\sum_{d=0}^{n-1}
 T_{n,k}^{(p)}(q_d^{(k)}).
 \label{eq:second-moment}
\end{equation}

Equation~\eqref{eq:second-moment} has a compact closed form for every
admissible $n,k,p$.  With $a=p^k$, $b=p^{n-k}$, define
\begin{equation}
 R_{n,k}^{(p)}=2N^3+8N^2+(N^2+4N)(a^2+b^2),
 \label{eq:trace-constant}
\end{equation}
and $S_\pm=\sum_{d=0}^{n-1}p^{\pm2q_d^{(k)}}$.  The sums evaluate to
\begin{align}
 S_+(n,k,p)&=p^{2k}+\frac{2(p^{2k}-p^2)}{p^2-1}\nonumber\\
 &\quad+n-2k+1,\\
 S_-(n,k,p)&=p^{-2k}+\frac{2(1-p^{-2(k-1)})}{p^2-1}\nonumber\\
 &\quad+n-2k+1.
 \label{eq:general-geometric-sums}
\end{align}
Direct substitution into Eq.~\eqref{eq:trace-polynomial} gives
\begin{equation}
 T_{n,k}^{(p)}(q)=R_{n,k}^{(p)}
 +2Na^2p^{-2q}+2b^2p^{2q}.
 \label{eq:general-trace-simplified}
\end{equation}
The overlap sequence contains $q=k$ once, each $1\le q<k$ twice, and $q=0$
$n-2k+1$ times; hence Eq.~\eqref{eq:general-geometric-sums} is exactly its two
geometric sums.  We therefore obtain
\begin{equation}
 \boxed{\begin{aligned}
 \E[(\Pplanark)^2]={}&\frac1{\rising N4}
 \left[R_{n,k}^{(p)}+\frac{2Na^2}{n}S_-\right.\\[-2pt]
 &\left.\hspace{15mm}+\frac{2b^2}{n}S_+\right]
 \end{aligned}}
 \label{eq:general-second-closed}
\end{equation}
and
\begin{equation}
 \boxed{\Var(\Pplanark)
 =\E[(\Pplanark)^2]-\left(\frac{a+b}{N+1}\right)^2.}
 \label{eq:general-variance}
\end{equation}
This formula is the main $k$-dependent result: the state-space integral has
been reduced to the elementary incidence profile of two intervals.

For the balanced qubit specialization $p=2$ and $k=r$, the geometric sums in
Appendix~\ref{app:second-moment} reduce Eq.~\eqref{eq:general-variance} to
\begin{equation}
 \boxed{
 \Var(\Pplanar)=
 \frac{\dfrac{20}{3n}(N^2-1)-8N}
 {(N+1)^2(N+2)(N+3)}}
 \quad (n\ \text{even}),
 \label{eq:variance-even}
\end{equation}
and
\begin{equation}
 \boxed{
 \Var(\Pplanar)=
 \frac{\dfrac{6}{n}(N^2-1)-9N}
 {(N+1)^2(N+2)(N+3)}}
 \quad (n\ \text{odd}).
 \label{eq:variance-odd}
\end{equation}
For even $n$, $C_{n,r}^{(2)}(d)=C_{n,r}^{(2)}(r-d)$ for $0\le d\le r$ and
$C_{n,r}^{(2)}(r)=C_{n,r}^{(2)}(0)$ because the interval at displacement $r$
is the complement of the original interval and has exactly the same purity.
Consequently,
\begin{equation}
 \Var(\Pplanar)\sim
 \begin{cases}
 \dfrac{20}{3nN^2},& n\ \text{even},\\[4pt]
 \dfrac{6}{nN^2},& n\ \text{odd}.
 \end{cases}
 \label{eq:variance-asymptotic}
\end{equation}
Since $n=\log_2N$, the variance is
$\Theta(N^{-2}/\log N)$ rather than a constant multiple of $N^{-2}$.

For even $n$, the known absolute-purity variance is~\cite{Facchi2010}
\begin{equation}
 \Var(\Pabsolute)=
 \frac{(N+1)f_2(N)-8N}{(N+1)^2(N+2)(N+3)},
 \label{eq:absolute-variance}
\end{equation}
where
\begin{align}
 f_2(N)={}&\frac{2^{r+1}}{\binom{2r}{r}}
 \sum_{j=0}^{r}\binom{r}{j}^2
 \left(4^{r/2-j}+4^{-(r/2-j)}\right).
 \label{eq:f2}
\end{align}
The relation to the present pair kernel is direct.  Averaging over one
representative of each complementary pair is identical, state by state, to
averaging over all $\binom{2r}{r}$ subsets.  For a fixed $r$-subset $A$, exactly
$\binom rd^2$ subsets $B$ satisfy $|A\setminus B|=d$, or equivalently
$|A\cap B|=r-d$.  Consequently,
\begin{equation}
 \E[(\Pabsolute)^2]
 =\frac{1}{\binom{2r}{r}\rising N4}
 \sum_{d=0}^{r}\binom rd^2 T_{2r,r}^{(2)}(r-d).
 \label{eq:absolute-from-kernel}
\end{equation}
Substitution of Eq.~\eqref{eq:trace-even-simplified} and subtraction of the
shared mean in Eq.~\eqref{eq:shared-mean} reproduce
Eqs.~\eqref{eq:absolute-variance} and \eqref{eq:f2}.  Thus the cyclic and
absolute calculations use the same two-subsystem Haar kernel but different
incidence multiplicities: one interval at each displacement versus a binomial
shell of balanced subsets.
Its asymptotic decay is approximately
$N^{-2.415}$~\cite{Facchi2010}, faster than the planar variance.  Thus, along
the even sizes for which Eq.~\eqref{eq:absolute-variance} applies, the ratio
$\Var(\Pplanar)/\Var(\Pabsolute)$ grows with system size even though both
functionals have the same mean.

\begin{figure*}[t]
 \centering
 \includegraphics[width=0.92\textwidth]{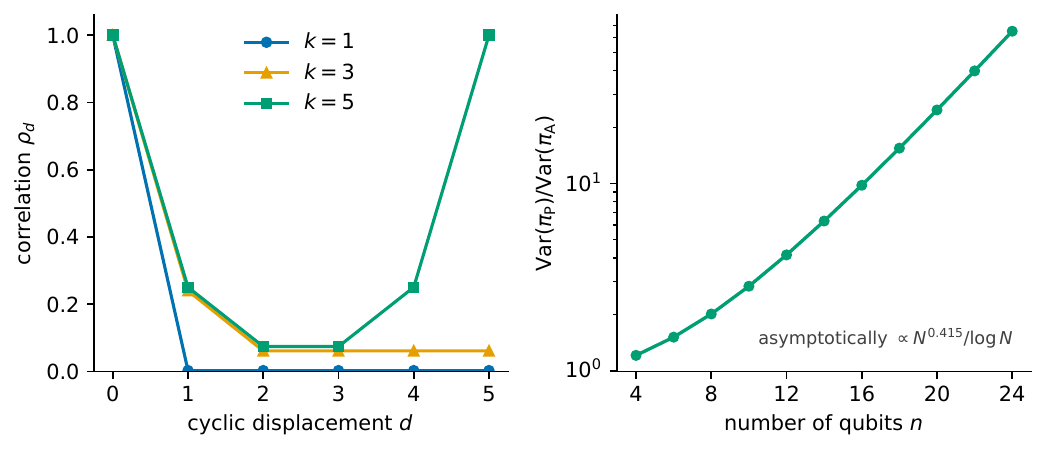}
 \caption{Geometry of planar-purity fluctuations.  Left: exact correlation
 coefficient $\rho_d=C_{10,k}^{(2)}(d)/C_{10,k}^{(2)}(0)$ for the nonbalanced
 choices $k=1,3$ and the balanced choice $k=5$.  The balanced endpoint
 $d=5$ is the complementary interval and obeys $\pi_A=\pi_{A^c}$ exactly;
 changing $k$ alters the complete displacement kernel.  Right: exact variance
 ratio between balanced planar and absolute purity for even $4\le n\le24$.
 The annotation gives the asymptotic growth implied by
 Eqs.~\eqref{eq:variance-asymptotic} and \eqref{eq:absolute-variance}; no fitted
 values enter either panel.}
 \label{fig:geometry}
\end{figure*}

Figure~\ref{fig:geometry} makes the origin of this difference explicit.  The
left panel normalizes $C_{n,r}^{(2)}(d)$ by the variance of one balanced-block
purity.  The curve is nonmonotone because an interval becomes progressively
closer to the complement of the original interval as $d$ approaches $r$; for a
pure state that complement has exactly the original purity.  The right panel
uses the exact finite-$n$ formulas and shows the growing
planar-to-absolute variance ratio.

\subsection{Higher moments}
\label{subsec:higher-moments}

Equation~\eqref{eq:haar-moment} also yields an exact finite representation of
every raw moment.  For a moment order $m$, let
$\bm s=(s_1,\ldots,s_m)$ and let $\tau_j^{(k)}(\bm s)\in S_{2m}$ be the product
of the disjoint transpositions $(2\ell-1,2\ell)$ for which site $j$ belongs to
$A_{s_\ell}^{(k)}$.  Then
\begin{equation}
 \E[(\Pplanark)^m]
 =\frac1{n^m\rising N{2m}}
 \sum_{\bm s\in\mathbb Z_n^m}
 \sum_{\sigma\in S_{2m}}
 p^{\sum_{j=0}^{n-1}c(\tau_j^{(k)}(\bm s)\sigma)},
 \label{eq:general-moment}
\end{equation}
where $c(\rho)$ is the number of cycles of $\rho$, including fixed points.
The product $\tau_j^{(k)}(\bm s)\sigma$ uses the convention stated below
Eq.~\eqref{eq:haar-moment}.  Equation~\eqref{eq:general-moment} follows by
expanding the $m$ copies of Eq.~\eqref{eq:planar-purity}, inserting one swap
for each purity, and factorizing the trace of every $U_\sigma$ site by site.
For $m=3$, the inner sum contains $6!=720$ replica permutations.  We evaluate
this exact finite sum with integer arithmetic at the system sizes considered
here; the resulting raw and centered third moments are rational numbers and are
provided in machine-readable form in the Supplemental Material.  For the
balanced qubit case we report the centered, standardized skewness
\begin{equation}
 \gamma_1=\frac{\E[(\Pplanar-\E\Pplanar)^3]}
 {\Var(\Pplanar)^{3/2}}
 \label{eq:skewness}
\end{equation}
rather than calling the unnormalized third cumulant itself the skewness.

\subsection{Balanced closed forms for arbitrary local dimension}
\label{subsec:qudits}

Specializing Eq.~\eqref{eq:general-variance} to $k=r=\lfloor n/2\rfloor$
gives particularly compact parity formulas for every $p\ge2$.  The mean is
\begin{equation}
 \E[\Pplanar]=\frac{p^r+p^{n-r}}{N+1}.
 \label{eq:qudit-mean}
\end{equation}
With $D_N=(N+1)^2(N+2)(N+3)$, the variance is
\begin{align}
 \Var(\Pplanar)_{\mathrm{even}}
 &=\frac{\dfrac{4(p^2+1)}{n(p^2-1)}(N^2-1)-8N}{D_N},
 \label{eq:qudit-variance-even}\\
 \Var(\Pplanar)_{\mathrm{odd}}
 &=\frac{\dfrac{2(p+1)}{n(p-1)}(N^2-1)
 -\dfrac{2(p+1)^2}{p}N}{D_N}.
 \label{eq:qudit-variance-odd}
\end{align}
The qubit formulas in Eqs.~\eqref{eq:variance-even} and
\eqref{eq:variance-odd} are recovered at $p=2$.  Appendix~\ref{app:qudits}
records the simplification, and the accompanying tests compare both these
balanced forms and the general-$k$ formula against direct overlap sums for
several values of $p$.

\subsection{Asymptotic regimes away from balance}
\label{subsec:general-asymptotics}

The closed form in Eq.~\eqref{eq:general-variance} separates three regimes at
fixed local dimension $p$.  If $k$ remains fixed as $n\to\infty$, let
\begin{equation}
 c_{k,p}=p^{2k}+\frac{2(p^{2k}-p^2)}{p^2-1}-2k+1.
 \label{eq:fixed-k-coefficient}
\end{equation}
Expansion of the exact rational expression gives
\begin{equation}
 \Var(\Pplanark)\sim
 \frac{2c_{k,p}}{n p^{2k}N^2}.
 \label{eq:fixed-k-asymptotic}
\end{equation}
If instead $k/n\to\alpha$ with $0<\alpha<1/2$, both sides of each cut grow and
the endpoint remains macroscopically separated from balance.  In that regime,
\begin{equation}
 \Var(\Pplanark)\sim
 \frac{2(p^2+1)}{(p^2-1)nN^2}.
 \label{eq:extensive-k-asymptotic}
\end{equation}
The balanced endpoint is nonuniform: complementary intervals coincide as
purity observables for even $n$, while odd $n$ retains $n$ distinct cuts.  From
Eqs.~\eqref{eq:qudit-variance-even} and
\eqref{eq:qudit-variance-odd}, its respective coefficients are
$4(p^2+1)/(p^2-1)$ and $2(p+1)/(p-1)$ instead of the coefficient in
Eq.~\eqref{eq:extensive-k-asymptotic}.

These regimes also sharpen the typical-versus-extremal comparison.  With
$\Delta_{k,p}=\E[\Pplanark]-p^{-k}$, Eq.~\eqref{eq:mean-gap} gives, for fixed
$k$,
\begin{equation}
 \frac{\Delta_{k,p}}{\sqrt{\Var(\Pplanark)}}
 \sim (p^{2k}-1)\sqrt{\frac{n}{2c_{k,p}}}.
 \label{eq:fixed-k-gap-ratio}
\end{equation}
For extensive $k$, the same ratio grows as $p^k\sqrt n$ up to a
$p$-dependent constant.  Thus exact planar uniformity becomes atypical in
standard-deviation units already at fixed $k$, and exponentially more so when
the interval size grows linearly with $n$.

\section{Numerical validation}
\label{sec:numerics}

\subsection{Sampling and estimators}

We tested the balanced qubit specialization $p=2$, $k=r$ with
$M=40{,}000$ independent Haar-random states for each $4\le n\le9$ and an
independent $M=120{,}000$-state run at $n=10$.  To test the scope of the general
formula directly, we also sampled the three block sizes $k=1,2,3$ on the same
$40{,}000$ six-qubit states and the two block sizes $k=1,2$ on the same
$40{,}000$ four-qutrit states.  A state vector was generated as
\begin{equation}
 \ket\psi=\frac{\bm g}{\|\bm g\|_2},
 \qquad g_j=X_j+iY_j,
 \label{eq:gaussian-haar}
\end{equation}
where the $X_j$ and $Y_j$ are independent standard normal variables.  The
rotational invariance of the complex Gaussian distribution makes
Eq.~\eqref{eq:gaussian-haar} exactly Haar distributed on the unit sphere; a
full Haar-random unitary is unnecessary.

For a subsystem $A$, the amplitude tensor was transposed so that the $A$
indices precede the complementary indices and then reshaped into a
$d_A\times d_B$ matrix $C_A$, with $d_A=p^{|A|}$ and $d_B=p^{n-|A|}$.  We evaluated
$\rho_A=C_AC_A^\dagger$ and $\pi_A=\Tr(\rho_A^2)$.  States were processed in
batches, so neither an $N\times N$ global density matrix nor all reduced
states were retained in memory.

Let $y_i=\pi_i-\bar\pi$ and
$\widehat m_j=M^{-1}\sum_i y_i^j$.  The reported variance is the unbiased estimator
\begin{equation}
 s^2=\frac{M}{M-1}\widehat m_2,
 \label{eq:sample-variance}
\end{equation}
and the reported sample skewness is the usual finite-sample corrected
estimator
\begin{equation}
 G_1=\frac{\sqrt{M(M-1)}}{M-2}
 \frac{\widehat m_3}{\widehat m_2^{3/2}}.
 \label{eq:sample-skewness}
\end{equation}
The theoretical skewness is obtained from
Eq.~\eqref{eq:general-moment} with $m=3$.  We estimate one-standard-error
uncertainties using $s/\sqrt M$ for the mean,
$[(\widehat m_4-\widehat m_2^2)/M]^{1/2}$ for the variance, and the empirical influence
function
\begin{align}
 I_i={}&\frac{y_i^3-\widehat m_3-3\widehat m_2y_i}
 {\widehat m_2^{3/2}}
 -\frac{3\widehat m_3(y_i^2-\widehat m_2)}
 {2\widehat m_2^{5/2}},\nonumber\\
 \operatorname{SE}(G_1)&\simeq
 \frac1M\left(\sum_{i=1}^M I_i^2\right)^{1/2}
 \label{eq:skewness-se}
\end{align}
for the skewness.  These are asymptotic Monte Carlo uncertainties; their
purpose is to quantify the finite-sample comparison, not uncertainty in the
exact formulas.

The balanced table uses legacy datasets from the original study at
$n=4,\ldots,8$.  Their pseudorandom states were not recorded in the preserved
notebook, so the archived arrays can be reanalyzed exactly but cannot be
regenerated bit for bit.  All later validation datasets are reproducible from
recorded seeds: 4909 for $n=9$, 4910 for the independent $n=10$ replication,
4606 and 4704 for the nonbalanced-qubit and qutrit checks, and 4104 and 4108
for the paired planar/absolute comparisons at $n=4$ and $8$.  The legacy paired
$n=10$ dataset, whose seed is also unavailable, is used only in
Fig.~\ref{fig:comparison} and is identified there.  For even $n$, the code
evaluates one member of each complementary interval pair because their purities
are algebraically identical; this gives the same $1/n$ average as duplicating
all $n$ terms.  Software versions, sample counts, seeds, and file checksums are
recorded in the Supplemental Material.

\begin{table*}[t]
\caption{Exact balanced planar-purity moments ($p=2$, $k=r$) and Monte Carlo
estimates.  The sample count is $40{,}000$ for $4\le n\le9$ and $120{,}000$ for
$n=10$.  Parentheses give one estimated standard error in the final quoted
digits.  The sample variance uses Eq.~\eqref{eq:sample-variance}; the displayed
exact skewness is a numerical evaluation of the six-replica finite sum in
Eq.~\eqref{eq:general-moment}.}
\label{tab:validation}
\begin{ruledtabular}
\begin{tabular}{c cc cc cc}
$n$ & $\E[\Pplanar]$ & sample mean
& $\Var(\Pplanar)$ & sample variance
& exact $\gamma_1$ & sample skewness\\
\hline
4  & 0.470588 & 0.470506(274) & $3.00492\times10^{-3}$ & $2.9974(247)\times10^{-3}$ & 0.667698 & 0.6548(156)\\
5  & 0.363636 & 0.363544(134) & $7.25050\times10^{-4}$ & $7.2357(637)\times10^{-4}$ & 0.744245 & 0.7529(195)\\
6  & 0.246154 & 0.246120(74)  & $2.16133\times10^{-4}$ & $2.17278(1751)\times10^{-4}$ & 0.525087 & 0.5346(163)\\
7  & 0.186047 & 0.185982(34)  & $4.54860\times10^{-5}$ & $4.5465(349)\times10^{-5}$ & 0.431626 & 0.4317(140)\\
8  & 0.124514 & 0.124475(17)  & $1.19099\times10^{-5}$ & $1.19966(889)\times10^{-5}$ & 0.275648 & 0.2926(134)\\
9  & 0.093567 & 0.0935818(78) & $2.44251\times10^{-6}$ & $2.444(18)\times10^{-6}$ & 0.210572 & 0.229(13)\\
10 & 0.062439 & 0.0624384(23) & $6.24056\times10^{-7}$ & $6.259(26)\times10^{-7}$ & 0.130969 & 0.1291(74)
\end{tabular}
\end{ruledtabular}
\end{table*}

\begin{table*}[t]
\caption{Independent fixed-seed tests of the general-$k$, general-$p$
formulas.  Each row uses $40{,}000$ Haar states; block sizes belonging to the
same $(n,p)$ pair were evaluated on the same states.  Here
$z=(\text{sample}-\text{exact})/\operatorname{SE}$, with the mean and variance
standard errors defined in the text.}
\label{tab:general-validation}
\begin{ruledtabular}
\begin{tabular}{ccc ccc ccc}
$p$ & $n$ & $k$ & exact mean & sample mean & $z_{\mu}$
& exact variance & sample variance & $z_{v}$\\
\hline
2 & 6 & 1 & 0.523077 & 0.523148 &  1.73 & $6.76017\times10^{-5}$ & $6.73444\times10^{-5}$ & $-0.44$\\
2 & 6 & 2 & 0.307692 & 0.307716 &  0.39 & $1.48531\times10^{-4}$ & $1.48283\times10^{-4}$ & $-0.20$\\
2 & 6 & 3 & 0.246154 & 0.246212 &  0.79 & $2.16133\times10^{-4}$ & $2.16093\times10^{-4}$ & $-0.02$\\
3 & 4 & 1 & 0.365854 & 0.365830 & $-0.54$ & $8.05465\times10^{-5}$ & $8.03060\times10^{-5}$ & $-0.35$\\
3 & 4 & 2 & 0.219512 & 0.219555 &  0.68 & $1.61093\times10^{-4}$ & $1.60052\times10^{-4}$ & $-0.82$
\end{tabular}
\end{ruledtabular}
\end{table*}

\subsection{Moment comparison}

\begin{figure*}[t]
 \centering
 \includegraphics[width=0.96\textwidth]{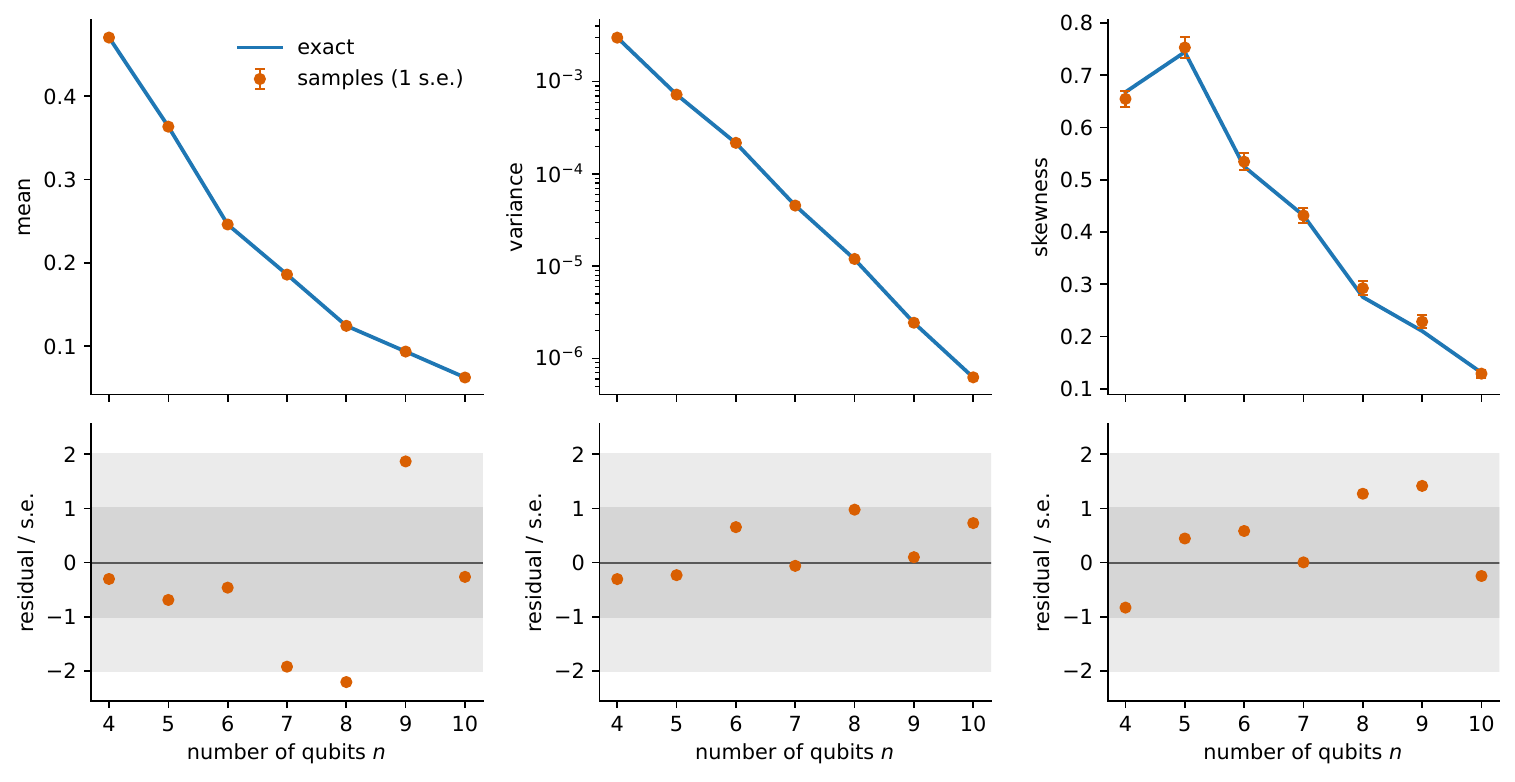}
 \caption{Balanced qubit planar-purity mean, variance, and standardized
 skewness for every $4\le n\le10$.  Blue curves are exact finite-$n$ values and
 connecting segments are guides to the eye; orange markers are Monte Carlo
 estimates from $40{,}000$ states through $n=9$ and $120{,}000$ states at
 $n=10$.  Error bars show one estimated standard error and may be smaller than
 the markers.  The lower row shows $(\text{sample}-\text{exact})/\operatorname{SE}$;
 dark and light bands mark one and two standard errors.  The variance panel is
 logarithmic.}
 \label{fig:moments}
\end{figure*}

Table~\ref{tab:general-validation} tests precisely the regimes not covered by a
balanced-qubit check.  All ten mean and variance residuals in that table lie
within $1.74$ estimated standard errors, including both nonbalanced qubit block
sizes and both qutrit block sizes.  Table~\ref{tab:validation} and
Fig.~\ref{fig:moments} give the complementary balanced test.  Across its 21
displayed moment comparisons, the mean, variance, and skewness residuals are
within $2.21$, $0.98$, and $1.42$ standard errors, respectively.  The odd
point at $n=9$ follows Eq.~\eqref{eq:variance-odd}, so no parity value is skipped
between $n=8$ and $10$.

The legacy $40{,}000$-state $n=10$ dataset gives sample variance
$6.3774(455)\times10^{-7}$, $3.00$ estimated standard errors above the exact
value.  We therefore generated an independent replication with
seed 4910 and three times as many states.  This $120{,}000$-state dataset,
reported in Table~\ref{tab:validation}, gives
$6.259(26)\times10^{-7}$, or $0.73$ standard errors above theory.  Both datasets
are archived.  The replication, the general-parameter simulations, the direct
overlap sums, and the exact all-subset recovery in
Eq.~\eqref{eq:absolute-from-kernel} provide independent checks of the
four-replica calculation.

\begin{figure}[t]
 \centering
 \includegraphics[width=\columnwidth]{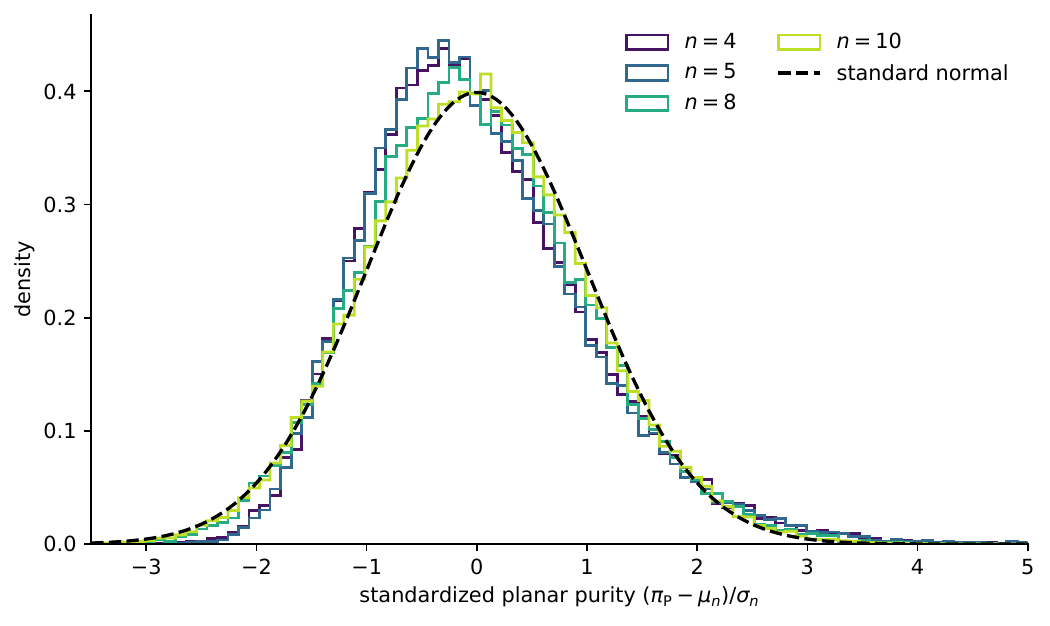}
 \caption{Balanced planar-purity histograms for representative Haar-random
 qubit sizes, using $40{,}000$ states for $n=4,5,8$ and $120{,}000$ for $n=10$,
 standardized with the exact mean and variance:
 $z=(\Pplanar-\E\Pplanar)/\sqrt{\Var(\Pplanar)}$.  The dashed curve is the
 standard normal density, not a fit.  Positive finite-size skewness appears as
 excess weight in the right tails; the curves do not by themselves establish a
 limiting Gaussian law.}
 \label{fig:standardized}
\end{figure}

Figure~\ref{fig:standardized} shows the standardized distributions together
with the standard normal density.  The exact skewness first increases from
$0.668$ at $n=4$ to $0.744$ at $n=5$ and then decreases over the remaining
sizes, reaching $0.131$ at $n=10$.  This finite-size pattern is not a
central-limit theorem.

\begin{figure*}[t]
 \centering
 \includegraphics[width=0.96\textwidth]{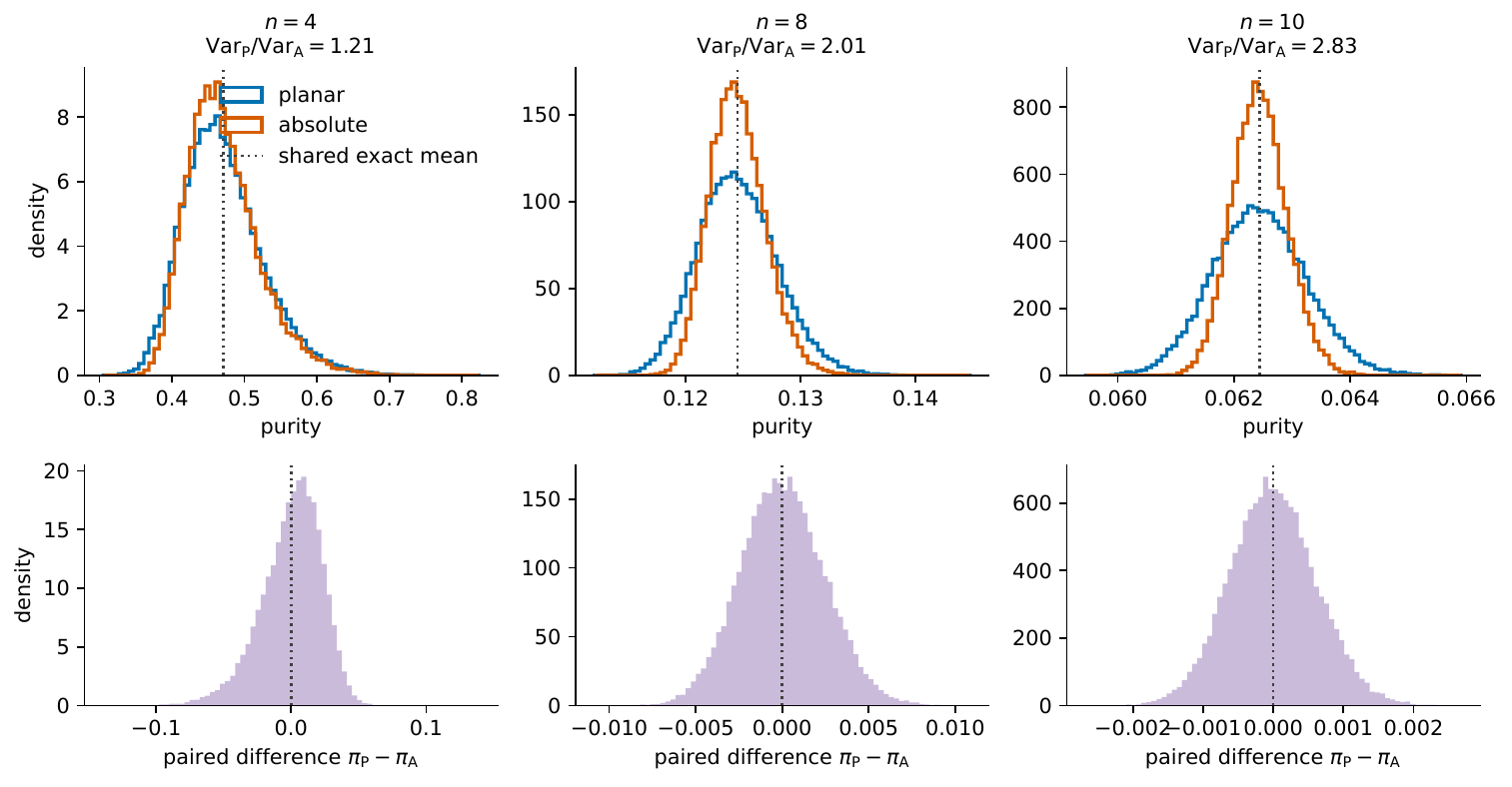}
 \caption{Balanced planar and absolute purity on paired Haar-random qubit
 states.  The top row shows marginal histograms and their common exact mean;
 each panel prints the exact variance ratio.  The bottom row uses the pairing
 directly by displaying $\Pplanar-\Pabsolute$, whose exact mean is zero.  Planar
 purity averages only contiguous cuts and is broader at all three sizes; the
 excess grows from $21\%$ at $n=4$ to a factor $2.83$ at $n=10$.  The $n=4,8$
 pairs use recorded seeds; the legacy $n=10$ paired array is retained with
 its provenance qualification stated in the text.}
 \label{fig:comparison}
\end{figure*}

Finally, Fig.~\ref{fig:comparison} compares planar and absolute purity on the
same Haar-random states for $n=4$ and $8$, and on the saved paired sample for
$n=10$.  The functionals have the same mean, while the planar distribution
becomes visibly broader.  The exact variance ratios are approximately $1.21$,
$2.01$, and $2.83$ for $n=4,8,10$, respectively.

\section{Discussion}
\label{sec:discussion}

\subsection{Typicality, geometry, and extremality}

The common mean in Eq.~\eqref{eq:shared-mean} is a consequence of Haar
invariance: every balanced subsystem of a fixed size has the same marginal
purity distribution.  Geometry first appears in a joint moment.  For the
planar functional, two summands are indexed by intervals and their covariance
depends only on the intersection size $q_d^{(k)}$.  For the absolute functional,
the average includes many more intersection patterns and many more cuts.
Equation~\eqref{eq:covariance-kernel} shows that the planar variance is not
merely the variance of one purity divided by the number of distinct intervals;
in the balanced examples displayed in Fig.~\ref{fig:geometry}, positive
cross-covariances supply a substantial part of the answer.

The general-$k$ result makes this geometric content quantitative.  The overlap
profile has $n-2k+1$ displacements with $q=0$, but only one with $q=k$ and two
for each intermediate value.  Increasing $k$ therefore changes both the
single-cut dimensions and the number of correlated interval pairs.  For
example, Eq.~\eqref{eq:general-variance} gives qubit variances
$1.4582\times10^{-7}$ at $k=1$ and $6.2406\times10^{-7}$ at $k=5$ for $n=10$.
This factor-$4.28$ difference is the exact combined consequence of subsystem
size and cyclic incidence.  The covariance
kernel consequently quantifies statistical redundancy under Haar measure:
intervals with large positive covariance contribute correlated rather than
independent fluctuations.  Whether the same intervals are redundant in an
optimization landscape depends on the ansatz and dynamics and is not decided
by the Haar calculation alone.

The parity dependence has an equally concrete origin.  With odd $n$, a
balanced block contains $(n-1)/2$ qubits and its complement has one additional
qubit, so all $n$ interval cuts are distinct.  With even $n$, the displacement
$r=n/2$ produces the complementary subsystem and exactly duplicates the
original purity.  In addition, a fixed balanced reduction has equal subsystem
and environment dimensions for even $n$, but dimension ratio $p$ for odd $n$.
The coefficients $20/3$ and $6$ in Eq.~\eqref{eq:variance-asymptotic}
therefore reflect both the different one-cut aspect ratios and the different
cyclic overlap multiplicities.

The result also separates typical entanglement from extremal entanglement.
For any $k$, the Haar mean lies above the planar $k$-uniform lower bound by
\begin{equation}
 \E[\Pplanark]-p^{-k}
 =\frac{p^{2k}-1}{p^k(N+1)}.
 \label{eq:mean-gap}
\end{equation}
For a balanced even system, this gap is of order $N^{-1/2}$, whereas the planar
standard deviation is of order $N^{-1}/\sqrt{\log N}$.  Thus the planar
$r$-uniform minimizer set is many standard deviations below the mean as $n$ grows.
A typical Haar state is highly entangled across each fixed balanced cut, but
exact simultaneous maximal mixing remains an extremal constraint.

\subsection{Operational scope and advantages over AME targets}

Planar uniformity retains a controlled subset of the resources provided by an
AME state.  Its first advantage is feasibility: minimal-support planar
$k$-uniform states exist for every $p\ge2$ and $n\ge2k$, whereas AME states are
absent for many party numbers and local dimensions~\cite{Wang2021,Huber2017}.
Its second advantage is geometric economy.  Certification requires $n$
translated $k$-site reductions---only $n/2$ distinct balanced purities for even
$n$---rather than all $\binom nk$ size-$k$ reductions (or one representative
of each complementary pair at balance).  This reduces the number of target
constraints from exponential to linear in the balanced regime, although the
measurement-shot complexity still depends on the chosen estimator.

The relaxation can also be expressed without making an assumption about a
search algorithm.  For a per-cut tolerance $\varepsilon\ge0$, let
\begin{align}
 \mathcal U_k(\varepsilon)&=\left\{\psi:
 \max_{|A|=k}\bigl[\pi_A(\psi)-p^{-k}\bigr]\le\varepsilon\right\},\nonumber\\
 \mathcal P_k(\varepsilon)&=\left\{\psi:
 \max_s\bigl[\pi_{A_s^{(k)}}(\psi)-p^{-k}\bigr]\le\varepsilon\right\}.
 \label{eq:approximate-set-inclusion}
\end{align}
Because the cyclic intervals are a subset of all $k$-site subsystems,
\begin{equation}
 \mathcal U_k(\varepsilon)\subseteq\mathcal P_k(\varepsilon).
 \label{eq:approximate-inclusion}
\end{equation}
Thus a planar target has a no-smaller feasible set under this uniform-tolerance
criterion.  At $k=r$, the left side is the corresponding approximate-AME
condition.  The inclusion supports the intuition that the planar constraint is
less restrictive, but it does not determine the Haar lower tail of the averaged
cost $\Pplanark$.

Several AME applications survive with a restricted access geometry.  If
$A=A_s^{(k)}$, then $\rho_A=I_A/p^k$ gives a Schmidt decomposition with
$p^k$ equal coefficients across $A|A^c$.  The state is therefore a maximally
entangled resource of dimension $p^k$ across every allowed contiguous cut and
supports standard teleportation across that cut~\cite{Bennett1993}.  The
bipartite entanglement resource used in the AME parallel-teleportation
construction~\cite{Helwig2012} is therefore available for sender--receiver
partitions whose smaller side is a contiguous block, subject to the same joint
operation assumptions.  What is lost is the freedom to choose an arbitrary
bipartition after distribution.

Wang's local-unitary orbit of a minimal-support planar $k$-uniform state forms
an orthogonal planar $k$-uniform basis~\cite{Wang2021}.  A dealer can encode a
classical secret in the basis label and distribute one subsystem to each party
around the ring.  Every contiguous coalition of at most $k$ parties then sees
the same maximally mixed state and is unauthorized, whereas all parties can
recover the label by a global basis measurement.  This is a
geometry-restricted data-hiding or classical secret-sharing primitive; it does
not assert recovery by every noncontiguous or larger coalition.  The
construction also connects naturally to $k$-uniform quantum
masking~\cite{Shi2021Masking}.  A full quantum-secret-sharing code additionally
requires off-diagonal decoupling conditions between codewords; planar
uniformity of the individual basis states alone is not sufficient.

Finally, at balance the state tensor is an isometry across every contiguous
half-system cut.  This is the ordered analogue of the perfect-tensor property
used in holographic codes.  Block-perfect tensors exploit precisely such a
relaxation to enlarge the available tensor and code families
~\cite{Harris2018,Steinberg2024}.  These existence, access-geometry, and
certification advantages make planar $k$-uniform states plausible laboratory
targets, while a claim of lower circuit depth requires a preparation model and
remains open.

\subsection{What the variance does not establish}

Both exact planar $k$-uniform and exact AME states form measure-zero subsets of
the continuous Haar ensemble.  To quantify how often an approximately planar
$k$-uniform state is encountered, one must specify a tolerance and study a
lower-tail probability such as
\begin{equation}
 \Pr\!\left[\Pplanark\le p^{-k}+\varepsilon\right].
 \label{eq:lower-tail}
\end{equation}
A larger variance at fixed mean is compatible with a broader lower tail but
does not determine Eq.~\eqref{eq:lower-tail}.  A distribution can have larger
variance because of its upper tail alone.  Large-deviation bounds, exact
cumulant control, importance sampling, or other rare-event methods are needed
before one can compare approximate planar-$k$-uniform and approximate-AME
sampling rates.

Likewise, planar $k$-uniformity concerns contiguous multipartite reductions.
It does not assert maximal pairwise entanglement between neighboring parties,
nor does AME imply large two-party entanglement for every pair.  Experimental
preparation claims additionally require a state family,
interaction graph, gate set, error model, and complexity measure.  None of
those operational assumptions is contained in a Haar variance.

\subsection{Finite-size shape}

After a small increase between $n=4$ and $n=5$, the exact skewness decreases
over the remaining sizes and reaches $0.131$ at $n=10$.  This is a
finite-size observation, not a proof of Gaussian convergence.  The variables
being averaged are neither independent nor drawn from an $n$-independent
distribution, and the number of distinct planar terms grows only linearly in
$n$.  Any central-limit theorem must therefore control the replica cumulants
and the changing covariance kernel.  Equation~\eqref{eq:general-moment}
provides an exact starting point but grows factorially in the replica number.

\section{Future directions}
\label{sec:future}

The exact pair kernel isolates the part of the problem that is universal under
Haar measure from the part that depends on subsystem geometry.  This
separation suggests several concrete extensions.

\subsection{Limiting laws and rare events}

A first problem is to determine the limiting distribution of standardized
planar purity.  Equation~\eqref{eq:general-moment} can be reorganized into
connected permutation diagrams, so a central-limit theorem would follow if
all standardized cumulants of order $m\ge3$ vanish.  Conversely, a surviving
connected family would identify a non-Gaussian limit.  The exact third
cumulant calculated here is only the first diagnostic; fourth and higher
cumulants must be controlled uniformly as the number of sites and the local
Hilbert-space dimension vary.  It would also be useful to determine whether
the fixed-$k$, extensive-$k$, and balanced limits belong to the same
universality class.

The more difficult companion problem is the lower tail in
Eq.~\eqref{eq:lower-tail}, which governs approximate planar-$k$-uniform search.
For a single bipartition, Coulomb-gas and saddle-point methods reveal phase
transitions in the purity distribution of random reduced states
~\cite{Giraud2007Distribution,Nadal2010}.  For planar purity, however, the
reduced states at different starts are correlated.  A useful goal is therefore
a large-deviation principle that retains the complete interval-overlap
profile, including possible changes of the optimizing eigenvalue
configuration as the tolerance is lowered.  Direct Haar sampling is poorly
suited to this regime; importance sampling, multicanonical methods, or a tilted
replica ensemble should be compared against exact small-system moments before
being extrapolated.

\subsection{Structured ensembles and entanglement dynamics}

Haar randomness may be replaced by experimentally or algorithmically
accessible ensembles.  Mean purity requires state moments through order two,
whereas variance and skewness require orders four and six.  State or unitary
designs thus provide a controlled hierarchy for deciding which results remain
exact and which acquire ensemble-dependent corrections
~\cite{Dankert2009,HarrowLow2009,Brandao2016}.  One specific problem is to find
the circuit depth at which the full displacement-resolved covariance
$C_{n,k}^{(p)}(d)$ approaches the Haar prediction, rather than testing only a
single cut or the mean.  Random local circuits exhibit nontrivial spatial
entanglement growth and fluctuation laws~\cite{Nahum2017}; planar purity would
probe how those local fluctuations become correlated when many translated
cuts are sampled simultaneously.

Random tensor networks offer a second structured setting.  Their entropies and
error-correcting properties are controlled jointly by tensor randomness and
network geometry~\cite{Pastawski2015,Hayden2016RandomTensor,Steinberg2024}.
Replacing the ring by boundary regions of a Euclidean or hyperbolic network
would test whether an incidence kernel analogous to
Eq.~\eqref{eq:covariance-kernel} can quantify fluctuations around minimal-cut
predictions.  Related questions include the effect of finite bond dimension,
nonuniform tensors, and approximate rather than perfect local building blocks.

\subsection{Subsystem design, inverse problems, and matching}

The subsystem family itself can be generalized.  Paths, higher-dimensional
tori, graph balls, unions of intervals, and hypergraph neighborhoods are
natural examples.  For a family $\mathcal F=\{A_\alpha\}$, define its binary
incidence matrix by
\begin{equation}
 M_{j\alpha}=\bm 1_{\{j\in A_\alpha\}},
 \qquad
 (M^{\mathsf T}M)_{\alpha\beta}=|A_\alpha\cap A_\beta|.
 \label{eq:general-incidence}
\end{equation}
For equal-size supports, the entries of this Gram matrix classify every pair
needed by the Haar second moment.  Families with unequal support sizes remain
tractable in principle, but require keeping the four region dimensions
separately instead of reducing them to the single overlap variable $q$.

This formulation also creates inverse and design problems.  One may ask which
incidence matrices minimize the Haar variance at fixed support size, which
families maximize sensitivity to a specified spatial correlation, or whether
the overlap spectrum can be reconstructed from measured purity covariances.
For a finite candidate list, assign the pair score
$w_{\alpha\beta}=g(|A_\alpha\cap A_\beta|)$ for a chosen design objective.
If each support may be used in at most one selected pair, maximizing the total
score is precisely a maximum-weight matching problem on the candidate graph.
Different selection rules lead to different combinatorial problems: pairwise
compatibility gives a clique problem, disjoint supports give set packing, and
higher-order compatibility gives a hypergraph problem.  Quantum query
algorithms for matching~\cite{AmbainisSpalek2006,Dorn2009} could accelerate
only this oracle-based outer optimization.  They neither prepare the target
state nor replace the Haar calculation.

\subsection{Preparation, measurement, and certification}

Variational circuits, stabilizer constructions, and tensor-network ansatzes
can be optimized against $\Pplanark-p^{-k}$, with explicit planar
$k$-uniform constructions supplying exact targets~\cite{Wang2021}.  A useful
preparation question is whether the reduced constraint family permits lower
depth on a geometrically local architecture than comparable AME targets; the
existence theorem alone does not answer it.  Certification is more immediate:
full tomography of every reduction is unnecessary.  Randomized measurements
already access R\'enyi entropies in many-body experiments~\cite{Brydges2019},
and classical shadows estimate many observables from shared data
~\cite{Huang2020,Elben2023}.  The translated intervals may therefore reuse
measurement settings and shots.

Because those interval estimators overlap, their sampling errors will be
correlated just as the intrinsic purities are.  A complete certification theory
should therefore derive the joint estimator covariance, optimize measurement
allocation over starts, include readout and gate noise, and construct a
finite-sample hypothesis test for
$\Pplanark\le p^{-k}+\varepsilon$.  Separating measurement noise from the Haar
state-to-state variance derived here is essential before using a measured
spread as evidence for or against approximate planar uniformity.

\subsection{Spatial stochastic systems and biophysical geometry}

Spatial population models provide a nonquantum setting in which neighborhood
incidence may organize correlations.  Moran dynamics on graphs depend on
update order and environmental heterogeneity~\cite{Moran1958,Kaveh2015,Kaveh2019};
the continuous-media extension treats weighted lattices while distinguishing
fecundity from viability~\cite{Gorgi2025Moran}.  If local replacement
neighborhoods are taken as supports $A_\alpha$, then
$M^{\mathsf T}M$ records how strongly those neighborhoods overlap.  A concrete
problem is to determine when this overlap is sufficient to classify two-point
fluctuations.  The covariance itself must be derived from the population
model's Markov generator, not from the Haar kernel.

The same incidence language applies to geometric ordering in bacterial
communities, where Voronoi tessellations predict spatial patterns
~\cite{Gorgi2026Geometric}.  Dirichlet descriptions and vertex/Voronoi models
relate cellular neighborhoods to packing, mechanics, and rigidity
~\cite{Honda1978,Farhadifar2007,Bi2015}.  Treating cells, adjacency stars, or
fixed-radius neighborhoods as supports would produce overlap spectra that can
be compared across seed distributions, boundaries, and growth rules.  This is
a transfer of geometric descriptors, not an identification of quantum purity
with a biological observable.

\section{Conclusion}
\label{sec:conclusion}

Planar $k$-purity is a faithful cost function whose attained minimum
characterizes planar $k$-uniform states.  Its two-interval Haar moment is the
same universal pair kernel that underlies absolute balanced purity; what
changes is the incidence multiplicity with which each overlap enters.  For
cyclic intervals this geometry can be summed exactly, giving the full
covariance kernel and closed variance for every admissible $k$ and $p$, with
distinct fixed-$k$, extensive, and balanced asymptotic regimes.  Independent
nonbalanced-qubit, qutrit, odd-$n$, and larger $n=10$ simulations support these
formulas, while the six-replica representation resolves finite-size skewness.

The central physical message is therefore sharper than a comparison of two
averages: Haar invariance makes every one-cut marginal statistic blind to where
a subsystem sits, whereas the variance of many cuts retains their pairwise
incidence profile.  Higher moments retain correspondingly higher-order
incidence information.  The relaxed planar constraint also exists throughout
the admissible parameter range and preserves communication and tensor-isometry
resources on contiguous cuts.  Lower-tail probabilities near the planar
$k$-uniform minimizer set and a limiting law for standardized planar purity
remain open.

\section*{Data and code availability}

The complete analysis code, regression tests, numerical samples, exact rational
third moments, machine-readable summary tables, figure-generation scripts, and
a checksum manifest are included in the accompanying reproducibility archive,
which is prepared as Supplemental Material and as an arXiv ancillary file.
Recorded seeds reproduce every validation dataset generated during the revised
analysis.  Legacy arrays whose seeds are unavailable are identified separately
and retained verbatim for reanalysis.

\appendix

\section{Balanced-qubit reduction of the second moment}
\label{app:second-moment}

For completeness, this appendix reduces the general result to the balanced
qubit case $p=2$, $k=r$.  Write $q_d=q_d^{(r)}$.  For even $n=2r$, the cyclic
overlap sequence for $d=0,\ldots,2r-1$ is
\begin{equation}
 q_d=|r-d|.
\end{equation}
Thus the values $1,\ldots,r-1$ occur twice, while $0$ and $r$ occur once.
For odd $n=2r+1$ and $d=0,\ldots,2r$, the sequence is
\begin{equation}
 q_d=
 \begin{cases}
 r-d,&0\le d\le r,\\
 d-r-1,&r+1\le d\le2r.
 \end{cases}
\end{equation}
Here $0,\ldots,r-1$ occur twice and $r$ occurs once.

For even $n$, substituting $x=w=2^q$ and $y=2^{r-q}$ into
Eq.~\eqref{eq:trace-polynomial} gives
\begin{equation}
 T_{2r,r}^{(2)}(q)=4N^3+16N^2+2N\,4^q+2N^2\,4^{-q}.
 \label{eq:trace-even-simplified}
\end{equation}
For odd $n$, $w=2^{q+1}$ and
\begin{equation}
 T_{2r+1,r}^{(2)}(q)=\frac92N^3+18N^2+4N\,4^q+N^2\,4^{-q}.
 \label{eq:trace-odd-simplified}
\end{equation}
Define $S_\pm=\sum_{d=0}^{n-1}4^{\pm q_d}$.  Direct evaluation of the two
finite geometric series gives
\begin{align}
 S_+^{\rm even}&=\frac53(N-1),
 &S_-^{\rm even}&=\frac53(1-N^{-1}),
 \label{eq:geometric-even}\\
 S_+^{\rm odd}&=\frac{5N-4}{6},
 &S_-^{\rm odd}&=\frac{8N-10}{3N}.
 \label{eq:geometric-odd}
\end{align}
Equations~\eqref{eq:trace-even-simplified}--\eqref{eq:geometric-odd} now
evaluate every term in Eq.~\eqref{eq:second-moment}; no asymptotic
approximation is used.  With
$D_N=(N+1)^2(N+2)(N+3)$, the resulting raw second moments are
\begin{align}
 \E[(\Pplanar)^2]_{\rm even}
 &=\frac{4N^3+20N^2+16N+\dfrac{20}{3n}(N^2-1)}{D_N},
 \label{eq:raw-second-even}\\
 \E[(\Pplanar)^2]_{\rm odd}
 &=\frac{\dfrac92N^3+\dfrac{45}2N^2+18N+\dfrac6n(N^2-1)}{D_N}.
 \label{eq:raw-second-odd}
\end{align}
Subtracting $[(2^r+2^{n-r})/(N+1)]^2$ gives the stated variances.  These formulas
also provide a direct numerical check on the permutation sum before any
asymptotic expansion is taken.

\section{Permutation audit}
\label{app:permutation-audit}

The entries in Table~\ref{tab:contraction-classes} can be reconstructed without an
index expansion.  On the four disjoint regions generated by
$A_s^{(k)}$ and $A_t^{(k)}$, the inserted replica permutations are, respectively,
\begin{equation}
 (12)(34),\qquad (12),\qquad (34),\qquad e.
\end{equation}
For each $\sigma\in S_4$, a region of Hilbert-space dimension $D$ contributes
$D^{c(\tau\sigma)}$.  Enumerating all 24 values of $\sigma$, collecting equal
monomials, and preserving their multiplicities yields
\begin{align}
\sum_{\sigma\in S_4}&x^{c((12)(34)\sigma)}
y^{c((12)\sigma)+c((34)\sigma)}w^{c(\sigma)}\nonumber\\
&=T_{n,k}^{(p)}(q),
\end{align}
with $x,y,w$ from Eq.~\eqref{eq:region-dimensions}.  The accompanying test
suite performs this 24-term enumeration independently for every admissible
block size and overlap through $n=10$ and checks the independently enumerated
sum against the collected polynomial.  It also verifies that the displacement average of
Eq.~\eqref{eq:covariance-kernel} reproduces both the general-$k$ result and the
balanced parity-dependent closed forms.

The six-replica implementation of Eq.~\eqref{eq:general-moment} uses the same
right-to-left composition convention.  For each ordered triple of interval
starts $(s_1,s_2,s_3)$, every site is assigned a three-bit membership mask.  The
mask selects a product of the disjoint swaps $(12)$, $(34)$, and $(56)$, and the
code sums $2^{c(\tau\sigma)}$ over all $\sigma\in S_6$ site by site.  All
contributions are accumulated as integers before division by
$n^3\rising N6$.  The centered third moment is then formed exactly as
$\E[(\Pplanar)^3]-3\E[(\Pplanar)^2]\E[\Pplanar]+2\E[\Pplanar]^3$ using rational
arithmetic.  The archive reports the numerator and denominator separately for
every $4\le n\le10$; a regression test also checks the resulting $n=4$
standardized skewness against an independently preserved value.

\section{Qudit displacement sums}
\label{app:qudits}

For local dimension $p$, the four region dimensions in
Eq.~\eqref{eq:region-dimensions} become
$x=p^q$, $y=p^{r-q}$, and $w=p^{n-2r+q}$.  The two simplified trace
polynomials are
\begin{align}
 T_{2r}^{(p)}(q)
 &=4N^3+16N^2+2N p^{2q}+2N^2p^{-2q},
 \label{eq:qudit-trace-even}\\
 T_{2r+1}^{(p)}(q)
 &=\frac{(p+1)^2}{p}N^3
 +\frac{4(p+1)^2}{p}N^2\nonumber\\
 &\quad+2pN p^{2q}+\frac{2}{p}N^2p^{-2q}.
 \label{eq:qudit-trace-odd}
\end{align}
The required displacement sums are
\begin{align}
 \sum_d p^{2q_d}\bigg|_{\rm even}
 &=\frac{p^2+1}{p^2-1}(N-1),\nonumber\\
 \sum_d p^{-2q_d}\bigg|_{\rm even}
 &=\frac{p^2+1}{p^2-1}(1-N^{-1}),
 \label{eq:qudit-sums-even}\\
 \sum_d p^{2q_d}\bigg|_{\rm odd}
 &=\frac{(p^2+1)N/p-2}{p^2-1},\nonumber\\
 \sum_d p^{-2q_d}\bigg|_{\rm odd}
 &=\frac{2p^2-(p^2+1)p/N}{p^2-1}.
 \label{eq:qudit-sums-odd}
\end{align}
Substitution into the analogue of Eq.~\eqref{eq:second-moment}, followed by
subtraction of Eq.~\eqref{eq:qudit-mean} squared, yields
Eqs.~\eqref{eq:qudit-variance-even} and
\eqref{eq:qudit-variance-odd}.

% Keep the expanded bibliography compact enough to avoid an almost-empty
% trailing page while retaining a conventional journal reference size.
\renewcommand{\bibfont}{\footnotesize}
\bibliography{references}

@article{Horodecki2009,
  author  = {Horodecki, Ryszard and Horodecki, Pawe\l{} and Horodecki, Micha\l{} and Horodecki, Karol},
  title   = {Quantum entanglement},
  journal = {Rev. Mod. Phys.},
  volume  = {81},
  pages   = {865--942},
  year    = {2009},
  doi     = {10.1103/RevModPhys.81.865}
}

@article{Facchi2006,
  author  = {Facchi, Paolo and Florio, Giuseppe and Pascazio, Saverio},
  title   = {Probability-density-function characterization of multipartite entanglement},
  journal = {Phys. Rev. A},
  volume  = {74},
  pages   = {042331},
  year    = {2006},
  doi     = {10.1103/PhysRevA.74.042331},
  eprint  = {quant-ph/0603281},
  archiveprefix = {arXiv}
}

@article{Facchi2010,
  author  = {Facchi, Paolo and Florio, Giuseppe and Marzolino, Ugo and Parisi, Giorgio and Pascazio, Saverio},
  title   = {Classical statistical mechanics approach to multipartite entanglement},
  journal = {J. Phys. A: Math. Theor.},
  volume  = {43},
  pages   = {225303},
  year    = {2010},
  doi     = {10.1088/1751-8113/43/22/225303},
  eprint  = {1002.2592},
  archiveprefix = {arXiv},
  primaryclass = {quant-ph}
}

@article{Scott2004,
  author  = {Scott, A. J.},
  title   = {Multipartite entanglement, quantum-error-correcting codes, and entangling power of quantum evolutions},
  journal = {Phys. Rev. A},
  volume  = {69},
  pages   = {052330},
  year    = {2004},
  doi     = {10.1103/PhysRevA.69.052330}
}

@article{Helwig2012,
  author  = {Helwig, Wolfram and Cui, Wei and Latorre, Jos\'{e} Ignacio and Riera, Arnau and Lo, Hoi-Kwong},
  title   = {Absolute maximal entanglement and quantum secret sharing},
  journal = {Phys. Rev. A},
  volume  = {86},
  pages   = {052335},
  year    = {2012},
  doi     = {10.1103/PhysRevA.86.052335},
  eprint  = {1204.2289},
  archiveprefix = {arXiv},
  primaryclass = {quant-ph}
}

@misc{Helwig2013,
  author  = {Helwig, Wolfram and Cui, Wei},
  title   = {Absolutely Maximally Entangled States: Existence and Applications},
  year    = {2013},
  eprint  = {1306.2536},
  archiveprefix = {arXiv},
  primaryclass = {quant-ph}
}

@article{Goyeneche2015,
  author  = {Goyeneche, Dardo and Alsina, Daniel and Latorre, Jos\'{e} I. and Riera, Arnau and {\.{Z}}yczkowski, Karol},
  title   = {Absolutely maximally entangled states, combinatorial designs, and multiunitary matrices},
  journal = {Phys. Rev. A},
  volume  = {92},
  pages   = {032316},
  year    = {2015},
  doi     = {10.1103/PhysRevA.92.032316},
  eprint  = {1506.08857},
  archiveprefix = {arXiv},
  primaryclass = {quant-ph}
}

@misc{Rajchel2025,
  author  = {Rajchel-Mieldzio\'{c}, Grzegorz and Bistro\'{n}, Rafa\l{} and Rico, Albert and Lakshminarayan, Arul and {\.{Z}}yczkowski, Karol},
  title   = {Absolutely maximally entangled pure states of multipartite quantum systems},
  year    = {2025},
  eprint  = {2508.04777},
  archiveprefix = {arXiv},
  primaryclass = {quant-ph}
}

@article{Wang2021,
  author  = {Wang, Yan-Ling},
  title   = {Planar $k$-uniform states: A generalization of planar maximally entangled states},
  journal = {Quantum Inf. Process.},
  volume  = {20},
  pages   = {271},
  year    = {2021},
  doi     = {10.1007/s11128-021-03204-y},
  eprint  = {2106.12209},
  archiveprefix = {arXiv},
  primaryclass = {quant-ph}
}

@article{Zyczkowski2001,
  author  = {{\.{Z}}yczkowski, Karol and Sommers, Hans-J\"urgen},
  title   = {Induced measures in the space of mixed quantum states},
  journal = {J. Phys. A: Math. Gen.},
  volume  = {34},
  pages   = {7111--7125},
  year    = {2001},
  doi     = {10.1088/0305-4470/34/35/335},
  eprint  = {quant-ph/0012101},
  archiveprefix = {arXiv}
}

@article{Collins2006,
  author  = {Collins, Beno\^{\i}t and {\'S}niady, Piotr},
  title   = {Integration with respect to the {Haar} measure on unitary, orthogonal and symplectic group},
  journal = {Commun. Math. Phys.},
  volume  = {264},
  pages   = {773--795},
  year    = {2006},
  doi     = {10.1007/s00220-006-1554-3},
  eprint  = {math-ph/0402073},
  archiveprefix = {arXiv}
}

@article{Lubkin1978,
  author  = {Lubkin, Elihu},
  title   = {Entropy of an $n$-system from its correlation with a $k$-reservoir},
  journal = {J. Math. Phys.},
  volume  = {19},
  pages   = {1028--1031},
  year    = {1978},
  doi     = {10.1063/1.523763}
}

@article{Page1993,
  author  = {Page, Don N.},
  title   = {Average entropy of a subsystem},
  journal = {Phys. Rev. Lett.},
  volume  = {71},
  pages   = {1291--1294},
  year    = {1993},
  doi     = {10.1103/PhysRevLett.71.1291},
  eprint  = {gr-qc/9305007},
  archiveprefix = {arXiv}
}

@article{Hayden2006,
  author  = {Hayden, Patrick and Leung, Debbie W. and Winter, Andreas},
  title   = {Aspects of generic entanglement},
  journal = {Commun. Math. Phys.},
  volume  = {265},
  pages   = {95--117},
  year    = {2006},
  doi     = {10.1007/s00220-006-1535-6},
  eprint  = {quant-ph/0407049},
  archiveprefix = {arXiv}
}

@article{Popescu2006,
  author  = {Popescu, Sandu and Short, Anthony J. and Winter, Andreas},
  title   = {Entanglement and the foundations of statistical mechanics},
  journal = {Nat. Phys.},
  volume  = {2},
  pages   = {754--758},
  year    = {2006},
  doi     = {10.1038/nphys444},
  eprint  = {quant-ph/0511225},
  archiveprefix = {arXiv}
}

@article{Goldstein2006,
  author  = {Goldstein, Sheldon and Lebowitz, Joel L. and Tumulka, Roderich and Zangh\`i, Nino},
  title   = {Canonical typicality},
  journal = {Phys. Rev. Lett.},
  volume  = {96},
  pages   = {050403},
  year    = {2006},
  doi     = {10.1103/PhysRevLett.96.050403},
  eprint  = {cond-mat/0511091},
  archiveprefix = {arXiv}
}

@article{Giraud2007Moments,
  author  = {Giraud, Olivier},
  title   = {Distribution of bipartite entanglement for random pure states},
  journal = {J. Phys. A: Math. Theor.},
  volume  = {40},
  pages   = {2793--2801},
  year    = {2007},
  doi     = {10.1088/1751-8113/40/11/014},
  eprint  = {quant-ph/0611285},
  archiveprefix = {arXiv}
}

@misc{Giraud2007Distribution,
  author  = {Giraud, Olivier},
  title   = {Purity distribution for bipartite random pure states},
  year    = {2007},
  eprint  = {0710.2045},
  archiveprefix = {arXiv},
  primaryclass = {quant-ph}
}

@article{Facchi2008,
  author  = {Facchi, Paolo and Florio, Giuseppe and Parisi, Giorgio and Pascazio, Saverio},
  title   = {Maximally multipartite entangled states},
  journal = {Phys. Rev. A},
  volume  = {77},
  pages   = {060304(R)},
  year    = {2008},
  doi     = {10.1103/PhysRevA.77.060304},
  eprint  = {0710.2868},
  archiveprefix = {arXiv},
  primaryclass = {quant-ph}
}

@article{Facchi2009,
  author  = {Facchi, Paolo and Florio, Giuseppe and Marzolino, Ugo and Parisi, Giorgio and Pascazio, Saverio},
  title   = {Statistical mechanics of multipartite entanglement},
  journal = {J. Phys. A: Math. Theor.},
  volume  = {42},
  pages   = {055304},
  year    = {2009},
  doi     = {10.1088/1751-8113/42/5/055304},
  eprint  = {0803.4498},
  archiveprefix = {arXiv},
  primaryclass = {quant-ph}
}

@article{Facchi2010Frustration,
  author  = {Facchi, Paolo and Florio, Giuseppe and Marzolino, Ugo and Parisi, Giorgio and Pascazio, Saverio},
  title   = {Multipartite entanglement and frustration},
  journal = {New J. Phys.},
  volume  = {12},
  pages   = {025015},
  year    = {2010},
  doi     = {10.1088/1367-2630/12/2/025015},
  eprint  = {0910.3134},
  archiveprefix = {arXiv},
  primaryclass = {quant-ph}
}

@article{Huber2017,
  author  = {Huber, Felix and G\"uhne, Otfried and Siewert, Jens},
  title   = {Absolutely maximally entangled states of seven qubits do not exist},
  journal = {Phys. Rev. Lett.},
  volume  = {118},
  pages   = {200502},
  year    = {2017},
  doi     = {10.1103/PhysRevLett.118.200502},
  eprint  = {1608.06228},
  archiveprefix = {arXiv},
  primaryclass = {quant-ph}
}

@article{Pastawski2015,
  author  = {Pastawski, Fernando and Yoshida, Beni and Harlow, Daniel and Preskill, John},
  title   = {Holographic quantum error-correcting codes: Toy models for the bulk/boundary correspondence},
  journal = {J. High Energy Phys.},
  volume  = {2015},
  number  = {6},
  pages   = {149},
  year    = {2015},
  doi     = {10.1007/JHEP06(2015)149},
  eprint  = {1503.06237},
  archiveprefix = {arXiv},
  primaryclass = {hep-th}
}

@misc{Steinberg2024,
  author  = {Steinberg, M. and Fan, J. and Harris, R. J. and Elkouss, D. and Feld, S. and Jahn, A.},
  title   = {Far from Perfect: Quantum Error Correction with (Hyperinvariant) Evenbly Codes},
  year    = {2024},
  eprint  = {2407.11926},
  archiveprefix = {arXiv},
  primaryclass = {quant-ph}
}

@misc{Geng2022,
  author  = {Geng, Runshi},
  title   = {Families of perfect tensors},
  year    = {2022},
  eprint  = {2211.15776},
  archiveprefix = {arXiv},
  primaryclass = {math.AG}
}

@misc{Gross2025,
  author  = {Gross, David and Goedicke, Paulina},
  title   = {Thirty-six officers, artisanally entangled},
  year    = {2025},
  eprint  = {2504.15401},
  archiveprefix = {arXiv},
  primaryclass = {quant-ph}
}

@misc{Trotta2026,
  author  = {Trotta, Giorgia and Scarafile, Paolo and Facchi, Paolo and
             Magnifico, Giuseppe and Mariano, Angelo and Parisi, Giorgio and
             Pascazio, Saverio and {\.{Z}}yczkowski, Karol},
  title   = {Multipartite entanglement of random states of qubits},
  year    = {2026},
  eprint  = {2605.10314},
  archiveprefix = {arXiv},
  primaryclass = {quant-ph}
}

@article{Dankert2009,
  author  = {Dankert, Christoph and Cleve, Richard and Emerson, Joseph and Livine, Etera},
  title   = {Exact and approximate unitary 2-designs and their application to fidelity estimation},
  journal = {Phys. Rev. A},
  volume  = {80},
  pages   = {012304},
  year    = {2009},
  doi     = {10.1103/PhysRevA.80.012304},
  eprint  = {quant-ph/0606161},
  archiveprefix = {arXiv}
}

@article{Brandao2016,
  author  = {Brand\~ao, Fernando G. S. L. and Harrow, Aram W. and Horodecki, Micha\l{}},
  title   = {Local random quantum circuits are approximate polynomial-designs},
  journal = {Commun. Math. Phys.},
  volume  = {346},
  pages   = {397--434},
  year    = {2016},
  doi     = {10.1007/s00220-016-2706-8},
  eprint  = {1208.0692},
  archiveprefix = {arXiv},
  primaryclass = {quant-ph}
}

@article{Elben2023,
  author  = {Elben, Andreas and Flammia, Steven T. and Huang, Hsin-Yuan and Kueng, Richard and Preskill, John and Vermersch, Beno\^\i{}t and Zoller, Peter},
  title   = {The randomized measurement toolbox},
  journal = {Nat. Rev. Phys.},
  volume  = {5},
  pages   = {9--24},
  year    = {2023},
  doi     = {10.1038/s42254-022-00535-2},
  eprint  = {2203.11374},
  archiveprefix = {arXiv},
  primaryclass = {quant-ph}
}

@misc{Gorgi2025Moran,
  author  = {Gorgi, Melika and Kaveh, Kamran and Aliakbarian, Navid and Ejtehadi, Mohammad Reza},
  title   = {Spatiotemporal {Moran} dynamics in continuous media},
  year    = {2025},
  eprint  = {2512.14171},
  archiveprefix = {arXiv},
  primaryclass = {q-bio.PE},
  doi     = {10.48550/arXiv.2512.14171}
}

@article{Gorgi2026Geometric,
  author  = {Gorgi, Melika and Kasallis, Summer J. and Trinh, Calvin and Ortiz de Ora, Lizett and Wiles, Travis J. and Siryaporn, Albert},
  title   = {Geometric ordering in bacterial communities},
  journal = {Proc. Natl. Acad. Sci. U.S.A.},
  volume  = {123},
  number  = {20},
  pages   = {e2526643123},
  year    = {2026},
  doi     = {10.1073/pnas.2526643123}
}

@article{Nadal2010,
  author  = {Nadal, C\'{e}line and Majumdar, Satya N. and Vergassola, Massimo},
  title   = {Phase transitions in the distribution of bipartite entanglement of a random pure state},
  journal = {Phys. Rev. Lett.},
  volume  = {104},
  pages   = {110501},
  year    = {2010},
  doi     = {10.1103/PhysRevLett.104.110501}
}

@article{HarrowLow2009,
  author  = {Harrow, Aram W. and Low, Richard A.},
  title   = {Random quantum circuits are approximate 2-designs},
  journal = {Commun. Math. Phys.},
  volume  = {291},
  pages   = {257--302},
  year    = {2009},
  doi     = {10.1007/s00220-009-0873-6},
  eprint  = {0802.1919},
  archiveprefix = {arXiv},
  primaryclass = {quant-ph}
}

@article{Nahum2017,
  author  = {Nahum, Adam and Ruhman, Jonathan and Vijay, Sagar and Haah, Jeongwan},
  title   = {Quantum entanglement growth under random unitary dynamics},
  journal = {Phys. Rev. X},
  volume  = {7},
  pages   = {031016},
  year    = {2017},
  doi     = {10.1103/PhysRevX.7.031016},
  eprint  = {1608.06950},
  archiveprefix = {arXiv},
  primaryclass = {cond-mat.stat-mech}
}

@article{Hayden2016RandomTensor,
  author  = {Hayden, Patrick and Nezami, Sepehr and Qi, Xiao-Liang and Thomas, Nathaniel and Walter, Michael and Yang, Zhao},
  title   = {Holographic duality from random tensor networks},
  journal = {J. High Energy Phys.},
  volume  = {2016},
  number  = {11},
  pages   = {009},
  year    = {2016},
  doi     = {10.1007/JHEP11(2016)009},
  eprint  = {1601.01694},
  archiveprefix = {arXiv},
  primaryclass = {hep-th}
}

@article{Brydges2019,
  author  = {Brydges, Tiff and Elben, Andreas and Jurcevic, Petar and Vermersch, Beno\^{\i}t and Maier, Christine and Lanyon, Ben P. and Zoller, Peter and Blatt, Rainer and Roos, Christian F.},
  title   = {Probing {R\'{e}nyi} entanglement entropy via randomized measurements},
  journal = {Science},
  volume  = {364},
  pages   = {260--263},
  year    = {2019},
  doi     = {10.1126/science.aau4963}
}

@article{Huang2020,
  author  = {Huang, Hsin-Yuan and Kueng, Richard and Preskill, John},
  title   = {Predicting many properties of a quantum system from very few measurements},
  journal = {Nat. Phys.},
  volume  = {16},
  pages   = {1050--1057},
  year    = {2020},
  doi     = {10.1038/s41567-020-0932-7},
  eprint  = {2002.08953},
  archiveprefix = {arXiv},
  primaryclass = {quant-ph}
}

@inproceedings{AmbainisSpalek2006,
  author    = {Ambainis, Andris and {\v{S}}palek, Robert},
  title     = {Quantum algorithms for matching and network flows},
  booktitle = {STACS 2006},
  series    = {Lecture Notes in Computer Science},
  volume    = {3884},
  pages     = {172--183},
  publisher = {Springer},
  year      = {2006},
  doi       = {10.1007/11672142_13},
  eprint    = {quant-ph/0508205},
  archiveprefix = {arXiv}
}

@article{Dorn2009,
  author  = {D{\"o}rn, Sebastian},
  title   = {Quantum algorithms for matching problems},
  journal = {Theory Comput. Syst.},
  volume  = {45},
  pages   = {613--628},
  year    = {2009},
  doi     = {10.1007/s00224-008-9118-x}
}

@article{Moran1958,
  author  = {Moran, P. A. P.},
  title   = {Random processes in genetics},
  journal = {Math. Proc. Cambridge Philos. Soc.},
  volume  = {54},
  number  = {1},
  pages   = {60--71},
  year    = {1958},
  doi     = {10.1017/S0305004100033193}
}

@article{Kaveh2015,
  author  = {Kaveh, Kamran and Komarova, Natalia L. and Kohandel, Mohammad},
  title   = {The duality of spatial death--birth and birth--death processes and limitations of the isothermal theorem},
  journal = {R. Soc. Open Sci.},
  volume  = {2},
  pages   = {140465},
  year    = {2015},
  doi     = {10.1098/rsos.140465}
}

@article{Kaveh2019,
  author  = {Kaveh, Kamran and McAvoy, Alex and Nowak, Martin A.},
  title   = {Environmental fitness heterogeneity in the {Moran} process},
  journal = {R. Soc. Open Sci.},
  volume  = {6},
  pages   = {181661},
  year    = {2019},
  doi     = {10.1098/rsos.181661}
}

@article{Honda1978,
  author  = {Honda, Hisao},
  title   = {Description of cellular patterns by {Dirichlet} domains: The two-dimensional case},
  journal = {J. Theor. Biol.},
  volume  = {72},
  number  = {3},
  pages   = {523--543},
  year    = {1978},
  doi     = {10.1016/0022-5193(78)90315-6}
}

@article{Farhadifar2007,
  author  = {Farhadifar, Reza and R{\"o}per, Jens-Christian and Aigouy, Beno\^{\i}t and Eaton, Suzanne and J{\"u}licher, Frank},
  title   = {The influence of cell mechanics, cell--cell interactions, and proliferation on epithelial packing},
  journal = {Curr. Biol.},
  volume  = {17},
  number  = {24},
  pages   = {2095--2104},
  year    = {2007},
  doi     = {10.1016/j.cub.2007.11.049}
}

@article{Bi2015,
  author  = {Bi, Dapeng and Lopez, J. H. and Schwarz, J. M. and Manning, M. Lisa},
  title   = {A density-independent rigidity transition in biological tissues},
  journal = {Nat. Phys.},
  volume  = {11},
  pages   = {1074--1079},
  year    = {2015},
  doi     = {10.1038/nphys3471}
}

@article{Bennett1993,
  author  = {Bennett, Charles H. and Brassard, Gilles and Cr\'{e}peau, Claude and Jozsa, Richard and Peres, Asher and Wootters, William K.},
  title   = {Teleporting an unknown quantum state via dual classical and {Einstein--Podolsky--Rosen} channels},
  journal = {Phys. Rev. Lett.},
  volume  = {70},
  pages   = {1895--1899},
  year    = {1993},
  doi     = {10.1103/PhysRevLett.70.1895}
}

@article{Shi2021Masking,
  author  = {Shi, Fei and Li, Mao-Sheng and Chen, Lin and Zhang, Xiande},
  title   = {$k$-uniform states and quantum information masking},
  journal = {Phys. Rev. A},
  volume  = {104},
  pages   = {032601},
  year    = {2021},
  doi     = {10.1103/PhysRevA.104.032601},
  eprint  = {2009.12497},
  archiveprefix = {arXiv},
  primaryclass = {quant-ph}
}

@article{Harris2018,
  author  = {Harris, Robert J. and McMahon, Nathan A. and Brennen, Gavin K. and Stace, Thomas M.},
  title   = {Calderbank--Shor--Steane holographic quantum error-correcting codes},
  journal = {Phys. Rev. A},
  volume  = {98},
  pages   = {052301},
  year    = {2018},
  doi     = {10.1103/PhysRevA.98.052301},
  eprint  = {1806.06472},
  archiveprefix = {arXiv},
  primaryclass = {quant-ph}
}

\end{document}